\documentclass[aps,prx,twocolumn,showpacs,superscriptaddress,longbibliography]{revtex4-2}  
\pdfoutput=1
\usepackage{graphicx}
\usepackage{amsmath,amssymb}
\usepackage{booktabs}
\usepackage{xcolor}
\usepackage{physics}
\usepackage[hidelinks]{hyperref}
\usepackage{supertabular}

\usepackage{empheq}
\usepackage{fancybox}

\usepackage[utf8]{inputenc}
\usepackage[tight,sf]{subfigure}%

\newcommand{\f}[2]{\frac{#1}{#2}}
\newcommand{\diff}{\mathrm{d}}
\renewcommand{\vec}[1]{{\bf #1}}
\newcommand{\parder}[2]{\frac{\partial #1}{\partial #2}}

\newcommand{\vk}{\vec{k}}
\newcommand{\vq}{\vec{q}}
\newcommand{\vp}{\vec{p}}
\newcommand{\vx}{\vec{x}}

\newcommand{\bs}{\mathbf}
\newcommand{\BigO}{\mathcal{O}}
\newcommand{\R}{\mathbb{R}}
\renewcommand{\Re}{\operatorname{Re}}

\newcommand{\T}{\mathbb{T}}
\newcommand{\C}{\mathbb{C}}
\newcommand{\Z}{\mathbb{Z}}

\newcommand{\menergy}{e}
\newcommand{\venergyd}{\varepsilon}
\newcommand{\mphase}{S}
\newcommand{\edens}{\mathcal{E}}

\newcommand{\ndensity}{\mathcal{N}}
\newcommand{\mdensity}{\mathcal{M}}
\newcommand{\totalenergy}{E}
\newcommand{\kcrit}{k_\text{c}}
\newcommand{\kmax}{k_\text{m}}

\begin{document}
	

\title{Emergent universal statistics in nonequilibrium systems\\ with dynamical scale selection }


\author{Vili Heinonen}
\affiliation{Department of Mathematics, Massachusetts Institute of Technology, 77 Massachusetts Avenue, Cambridge,~MA~02139, USA}
\affiliation{Department of Mathematics and Statistics, University of Helsinki, P.O. Box 68, FI-00014 Helsingin yliopisto, Finland}

\author{Abel J. Abraham}
\affiliation{
	Department of Mathematics,
	University of North Carolina at Chapel Hill,
	120 E Cameron Avenue,
	Chapel Hill, NC 27599, USA}

\author{Jonasz S\l{}omka}
\affiliation{Institute of Environmental Engineering, Department of Civil, Environmental, and Geomatic Engineering, ETH Zurich, 08093 Zurich, Switzerland}

\author{Keaton J. Burns}
\affiliation{Department of Mathematics, Massachusetts Institute of Technology, 77 Massachusetts Avenue, Cambridge,~MA~02139, USA}
	
\author{Pedro J. S\'aenz}
\affiliation{
	Department of Mathematics,
	University of North Carolina at Chapel Hill,
	120 E Cameron Avenue,
	Chapel Hill, NC 27599, USA}
	
\author{J\"orn Dunkel} 
\affiliation{Department of Mathematics, Massachusetts Institute of Technology, 77 Massachusetts Avenue, Cambridge,~MA~02139, USA}
\email{dunkel@mit.edu}

\date{\today}

\begin{abstract}
Pattern-forming nonequilibrium systems are ubiquitous in nature, from driven quantum matter and biological life forms to atmospheric and interstellar gases. Identifying universal aspects of their far-from-equilibrium dynamics and statistics poses major conceptual and practical challenges due to the absence of energy and momentum conservation laws. Here, we experimentally and theoretically investigate the statistics of prototypical nonequilibrium systems in which inherent length-scale selection confines the dynamics near a mean energy hypersurface. Guided by spectral analysis of the field modes and scaling arguments, we derive a universal nonequilibrium distribution for kinetic field observables. We confirm the predicted energy distributions in experimental observations of Faraday surface waves, and in random scattering and active turbulence simulations. Our results indicate that pattern dynamics and transport in driven physical and biological matter can often be described through monochromatic random fields, suggesting a path towards a unified statistical field theory of nonequilibrium systems with length-scale selection.
\end{abstract}

\maketitle


\section{Introduction}
Nonequilibrium systems balance energy uptake and dissipation to create complex dynamical structures across a wide range of length and time scales \cite{CrossHohenberg_RMP}, from turbulent flows \cite{Cardy2008}  in driven quantum \cite{2014Barenghi_PNAS} and classical \cite{2009Toschi_AnnuRev} fluids to the self-organized vortex patterns in active suspensions \cite{2020Duclos_Science,2012Wensink}. Recent major advances in the theoretical \cite{Ruelle2014,Goldenfeld2017,Li2019} and computational \cite{2020Gompper_NatRevPhys} modeling of nonequilibrium pattern-forming phenomena \cite{2002Aronson_RMP,2013Marchetti_Review} have led to a substantially improved understanding of transport processes in physics \cite{1998Dittrich_QTransport,2009Toschi_AnnuRev} and biology \cite{2009HanggiMarchesoni_RMP,2012TaylorStocker_Science}.  Despite such progress, however, there currently exists no unifying statistical field theory for far-from-equilibrium systems \cite{Oettinger2005,Seifert2012,Gnesotto2018,Li2019,Golestanian2019} on par with equilibrium thermodynamics. Perhaps the most fundamental difficulty in identifying universal nonequilibrium  statistical principles lies in the fact that systems can be driven out of equilibrium in various ways \cite{CrossHohenberg_RMP}.  Among the most widely studied driving mechanisms are temperature \cite{CrossHohenberg_RMP,Li2019,2010Lohse_AnnuRev} and pressure \cite{hino_1976_jfm} gradients, mechanical \cite{Ciliberto1985,1996KUDROLLI,2008Corte_NatPhys} or electromagnetic \cite{2011Kelley_NatPhys} forcing protocols, and chemical reservoirs \cite{1998Ebeling_PRL,2005Golestanian_PRL,2012PawelReview}. This diversity, combined with the absence of conservation laws, makes it challenging to find common statistical descriptions for these systems.

\begin{figure}[!b]
	\includegraphics[width = \columnwidth]
	{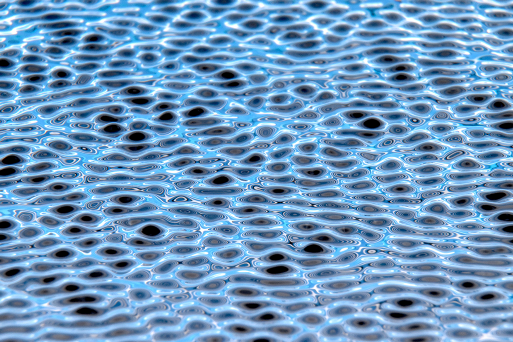}
	\caption{\textbf{
	Weakly chaotic Faraday surface waves exhibiting scar patterns.}
	Faraday waves  emerging on a vertically oscillating bath of water (Movies~1 and 2, Sec.~\ref{sec:methods}) are representatives of a broader class of non-equilibrium systems with spontaneous scale selection.  The photograph shows an oblique view of the fluid surface \cite{2017Harris}.  The shadows cast by larger wave crests give rise to dark scar-like patterns corresponding to regions of higher surface-gradient energy. In our experiments, the dynamically evolving surface height fields were reconstructed using a free-surface Schlieren technique \cite{Wildeman2018} (Sec.~\ref{sec:methods}). Faraday wavelength:  {$\lambda_\text{F}\approx 4.6$\,mm} (Sec.~\ref{sec:methods}).}
	\label{fig:faraday_waves}
\end{figure}

\begin{figure*}
	\includegraphics[width = 1.97\columnwidth]{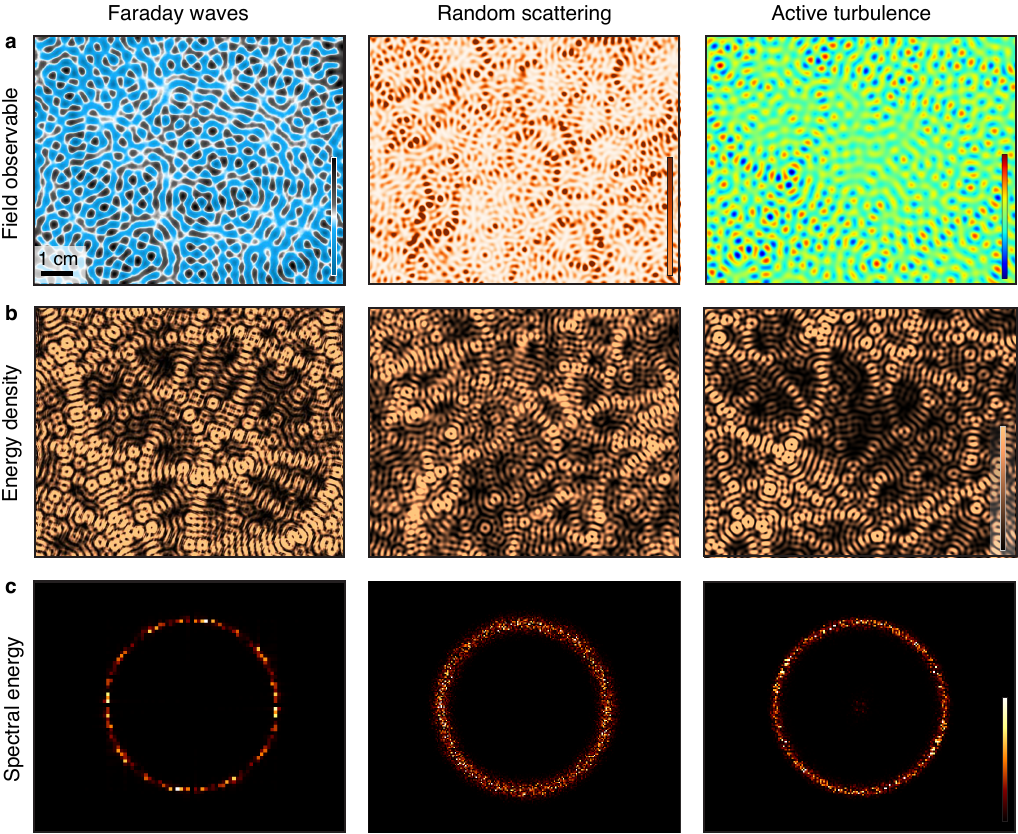}
	\caption{\textbf{Non-equilibrium field dynamics with length scale selection in experiment and simulations.} 
	\textbf{a},~Snapshots from our Faraday wave experiments (Fig.~\ref{fig:faraday_waves} and Movie~2), random scattering simulations in a smooth isotropic random potential (Eq.~\eqref{eq:schrodinger1}, Movie~3), and active turbulence simulations (Eq.~\eqref{activeNS-a}, Movie~4). Colors indicate the normalized height field $h$, number density $|\psi|^2$ and vorticity field, respectively.
	\textbf{b},~Associated {real space} energy densities {(Eq.~\eqref{energy})} reveal qualitatively similar structures across the different systems: the surface gradient energy of the Faraday waves, and the kinetic energies of the quantum and active fluid systems are characterized by scars extending throughout the system. 
	\textbf{c},~
	Spectral energy { $\menergy_\vec{k}$ (see Eq.\eqref{Fenergy}) at modes $\vec{k}$} shows that the energy of the system is concentrated within a narrow shell of  a fixed wave-number radius.
	Each panel represents a typical snapshot of the dynamical system at a time much larger than the initial relaxation period. See Sec.~\ref{sec:methods} for details and parameters of experiments, simulations and colorbar limits. }
	\label{fig:main_figure}
\end{figure*}

\par
Here, we  report progress on this longstanding problem by focusing on pattern-forming systems in which an intrinsic length-scale selection mechanism \cite{CrossHohenberg_RMP} effectively reduces the number of microscopic degrees of freedom. This large subclass of nonequilibrium systems comprises a broad  spectrum of physically and biologically important phenomena, including hydrodynamic and elastic instabilities \cite{Swift1977,1989Gollub_PRL,Ciliberto1985,Douady1990,Kumar1994,Zhang1995,1996KUDROLLI,Kumar1996,Kahouadji2015,Stoop2015,2017Lee}, liquid--solid phase transitions in quantum superfluids \cite{Pomeau1994,Yoo2010,Heinonen2019}  and active turbulence \cite{2012Sokolov,2012Wensink,Bratanov08122015,2018Yeomans_NatComm}. By combining experiments, theory and large-scale simulations, we  demonstrate that the competition between  length-scale selection and nonlinear mode-mixing can lead to the emergence of universal superstatistics \cite{Beck2005} for the relevant energetic field observables. Specifically, our results show that periodically forced Faraday waves \cite{1989Gollub_PRL,1996KUDROLLI} on the surface of water (Fig.~\ref{fig:faraday_waves}, Movies~1 and 2), quantum scattering~(Movie~3), and active turbulence~(Movie~4) can be jointly described through monochromatic random fields. Building on this insight, we illustrate the practical potential for modeling  nonequilibrium transport processes by constructing a generalized Langevin dynamics for passive tracer particles advected by a dense active microbial suspension.

\par
The unified statistical description of the three vastly different nonequilibrium systems (Fig.~\ref{fig:main_figure}a) investigated here becomes possible because, in each case, self-organized  dynamical length-scale selection (Fig.~\ref{fig:main_figure}b) concentrates the mode energies in a narrow circular shell in Fourier space (Fig.~\ref{fig:main_figure}c). In position space, the distinct  scale selection and mode interaction mechanisms governing Faraday waves, random scattering and active turbulence manifest themselves as visually similar, dynamically evolving scar-like structures \cite{1987Heller_PRL} in the local energy densities (Fig.~\ref{fig:main_figure}b). Our combined experimental, theoretical and numerical analysis  suggests that a wide range of nonequilibrium systems displaying such a phenomenology can be described by the same universal energy statistics.

\section{Model systems}
\subsection{Faraday waves}
In our experiments, we studied the dynamics of Faraday surface waves \cite{1989Gollub_PRL,1996KUDROLLI,Douady1990} on a vertically vibrated water bath (Movie~1).
In this parametrically-excited system, the energy injected by the external periodic forcing is balanced by internal viscous dissipation \cite{Kumar1994}. Owing to the oscillatory nature of the effective gravitational acceleration $G(t) = g + D \cos(\omega t)$ acting on the fluid, the flat free surface becomes unstable to subharmonic waves when the driving amplitude $D$ exceeds the critical Faraday threshold $D_\text{F}$. At  threshold, the competition between gravity and surface tension $\sigma$ results in a preferred Faraday wavelength~$\lambda_\text{F}=2\pi/k_\text{F}{=2\pi/k_\text{c}}$~dictated by the standard {capillary-gravity} dispersion relation  $(\omega/2)^2 =  g k_\text{F} + \sigma k_\text{F}^3/\rho$, where $\rho$ denotes the liquid density \cite{Kumar1994} (Sec.~\ref{sec:methods}). Our statistical analysis below is based on 3D surface measurements (Movie~1) of weakly nonlinear Faraday waves emerging in the super-critical driving regime \cite{1989Gollub_PRL} in which the scarred wave patterns evolve chaotically while maintaining a dominant Faraday wavelength~{$\lambda_\text{F}\approx 4.6$\,mm} (Fig.~\ref{fig:main_figure}a and Movie~2; Sec.~\ref{sec:methods}). 

\subsection{Random scattering} 
Dynamically evolving scarred patterns \cite{1984Heller,1987Heller_PRL} similar to those observed in Faraday waves arise in 2D random scattering \cite{Plisson2013} described by the {Schr\"odinger equation (Movie~3) 
\begin{align}\label{eq:schrodinger1}
i\hbar \partial_t\psi(t,\vx) =\left(\frac{-\hbar^2 \nabla^2}{2m}+V(\vx)\right)\psi(t,\vx)
\end{align}
where  $\psi(t,\vec{x})$ is the wave function, and the random potential $ V(\vec{x}) $  facilitates energy exchange between} the momentum modes~$e^{i\vec{k}\cdot\vec{x}}$.  We simulated Eq.~(\ref{eq:schrodinger1}) on a large periodic domain of size $L\times L$ with an initial uniform plane wave with momentum~$\hbar\vec{k}_\text{c}$. After a relaxation period $t_\text{rs}\sim 0.8$\,$m L^2/\hbar$ (Sec.~\ref{sec:methods}), the quantum state became isotropic  due to quasi-elastic wave scattering \cite{Plisson2013} by the smooth random  potential $ V $ (Fig.~\ref{fig:main_figure}; Movie~3) with a long-range Gaussian correlation function with zero mean, and small variance {compared to the initial wavepacket momentum} (Fig.~\ref{fig:potential}). Therefore, the dominant pattern length-scale (Fig.~\ref{fig:main_figure}b-c) is determined by~$k_\text{c}$, but not by the external scattering  potential~$V$ (Sec.~\ref{sec:methods}) which acts as an energy reservoir for the momentum modes. 

\subsection{Active turbulence}
Another entirely different nonequilibrium process exhibiting similar  energy localization in  Fourier space is active turbulence, as seen in dense bacterial  suspensions \cite{2012Sokolov, 2012Wensink,Bratanov08122015}. This widely studied phenomenon belongs to a broader class of linearly forced fluid flows, which also encompasses nonlinear seismic wave propagation \cite{1993BeNi_PhysD} and  soft-mode turbulence \cite{PhysRevE.77.035202} (Nikolaevskiy chaos). An effective phenomenological description of such pattern-forming flows is given by the linearly forced Navier-Stokes equations \cite{1993BeNi_PhysD,PhysRevE.77.035202,Slomka2017a} (Movie~4)
\begin{subequations}\label{activeNS-a}
\label{activeNS}
\begin{align}
\partial_t \vec{v} + \vec{v}\cdot \nabla \vec{v} &= -\nabla p + \nabla \cdot \boldsymbol{\sigma}, \\
\nabla \cdot \vec{v} &= 0,
\end{align}
\end{subequations}
where $p(t,\vec{x})$ is the local pressure, and the 2D incompressible velocity field $\vec {v}(t,\vec{x})=(\partial_y \psi,-\partial_x \psi)$ is determined by its stream function $\psi(t,\vec{x})$. The  phenomenological stress tensor {$\boldsymbol{\sigma}(t,\vec{x}) = (\Gamma_0 - \Gamma_2 \nabla^2 + \Gamma_4 \nabla^4)[\nabla \vec{v} + (\nabla \vec{v})^T  ]$ with $\Gamma_{0},\Gamma_{4} > 0 $} and $\Gamma_2 <0$ accounts for large- and small-scale  dissipation and intermediate-scale energy injection from an active component, such as swimming microbes, into the fluid medium \cite{1993BeNi_PhysD,PhysRevE.77.035202,Slomka2017a}. Energy transfer from active to dissipative modes occurs through the advective nonlinearity in Eq.~\eqref{activeNS-a}. In qualitative agreement with experimental observations for bacterial suspensions \cite{2012Sokolov, 2012Wensink,Slomka2017a},  Eqs.~\eqref{activeNS} predict flows with a typical vortex size $\Lambda = \pi\sqrt{2\Gamma_4/(-\Gamma_2)} = \pi/k_\text{c}$ that exhibit scar patterns in the  vorticity field $\omega = \nabla \times \vec{v}$ (Fig.~\ref{fig:main_figure}a right;  Movie~4).  
\par

\begin{figure*}
	\includegraphics[width = 2.\columnwidth]{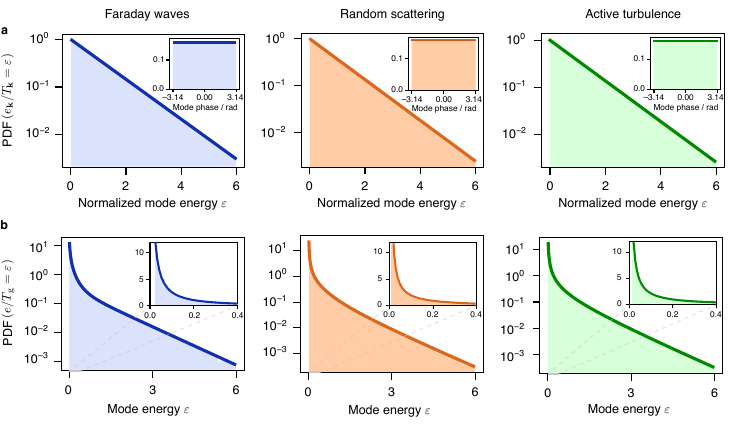}
	\caption{\textbf{Emergent universal statistics in experiment and simulations.} 
	\textbf{a}, Energy distribution functions for a representative subset of individual energy modes follow exponential Boltzmann distributions with different mode temperatures. The figures show the statistics for individual modes normalized by the mode temperature $T_\vec{k} = \langle {\menergy_\vec{k}} \rangle_t$. Insets: 
	all systems show uniform statistics for phases of Fourier modes.
	See Sec.~\ref{sec:methods} for details and parameters of experiments and simulations.
	\textbf{b}, Probability density functions  (PDFs) of the Fourier mode energies ${\menergy_\vec{k}}$ measured in the experiment (blue) and simulations (orange, green) follow the predicted superstatistics (solid lines) given by Eq.~\eqref{eq:superstatistics} with system specific global temperatures {$T_\text{g}$ (Sec.~\ref{sec:methods})}. 
	We introduce a low-energy cutoff allowing for direct normalization of $\ndensity$ (see Sec.~\ref{sec:methods}). {Here $\ndensity \propto \text{PDF}(e/T_\text{g} = \varepsilon)$}.
	Insets: blow-ups of the energy statistics at low energies shown on a linear scale.
	See also Figs.~\ref{fig:autocorrelation}-\ref{fig:divergence} for additional analysis of the energy distributions.}
	\label{fig:statistics}
\end{figure*}

\section{Unified description}
Despite their fundamental physical differences and nonequilibrium nature, we found that Faraday waves, 2D random scattering and active turbulence are representatives of a joint statistical  class whose essential energetic properties can be predicted from basic energy and symmetry considerations. To make this statement precise, we start by noting that all three systems possess a field-deformation energy of the form
\begin{equation}\label{energy}
	\totalenergy(t) \propto \int \diff \vec{x}\, 
	 \frac{1}{2}\,|\nabla \psi{(t,\vec{x})}|^2.
\end{equation}
For Faraday waves, $\psi$ is the surface height field and ${\totalenergy}$ the surface energy. In random scattering, ${\totalenergy}$ is the kinetic energy of the complex-valued wave function  $\psi$. In active turbulence, $\psi$ is the real-valued stream function and ${\totalenergy}$ is the kinetic energy of the suspension. In terms of the wave vectors $\vec{k}$ and Fourier amplitudes $\hat{\psi}_{\vec{k}}(t)$, the energy~\eqref{energy} can be expressed as
\begin{equation}\label{Fenergy}
	{\totalenergy}(t) = {C_\totalenergy} A\sum_{\vec{k}}  \frac{k^2}{2} |\hat{\psi}_{\vec{k}}(t) |^2 =: \sum_{\vec{k}} {\menergy_\vec{k}(t)},
\end{equation}
where ${A=L^2}$ is the {area} of the system, ${C_\totalenergy}$ is a system-specific constant (Sec.~\ref{sec:methods}), {and $\menergy_\vec{k}= C_\totalenergy A (k^2/2)|\hat{\psi}_{\vec{k}}|^2$ is the spectral energy of mode $\vec{k}$}. 
\par
We now formulate three
 general key criteria that enable statistical predictions for pattern-forming nonequilibrium systems with energies of the form~\eqref{Fenergy}:
\begin{enumerate}

\item{
The dynamics are ergodic meaning that, after a system dependent relaxation time, the system reaches a statistically stationary state that does not depend on the initial state. (Figs.~\ref{fig:main_figure}c).}

\item {
The steady state is described by a Gaussian field $\hat \psi_\Vec{k}$, whose correlations decay fast in time compared to observation time scales.}

\item {
The length scale selection mechanism localizes the energy on a narrow isotropic ring in Fourier space.
}
\end{enumerate}

\par
All three criteria are satisfied by our three example systems: The energies at modes $\vec{k}$ occupy the energy shell isotropically regardless of the details of the initial condition (Fig.~\ref{fig:main_figure}c and Figs.~{\ref{fig:autocorrelation}-\ref{fig:radial_energy distribution}}). We also calculated the autocorrelation and pair correlation functions for mode energies $\menergy_\vec{k}$ and verified that correlations decay exponentially (App.~\ref{sec:correlations}). {Beyond this numerical evidence, we argue that the Gaussian nature of the field (criterion 2) may in part follow from the system dynamics being confined to a narrow active ring (criterion 3). When the dynamics are confined to an active ring, the high-order cumulants can be expected to decay quickly in the active turbulence case because the nonlinear coupling between modes is small (App.~\ref{sec:cumulants}). This stands in contrast to passive turbulence, where the statistics are known to be strongly non-Gaussian in the dissipative range \cite{li2005}.}

The Gaussian distribution of $\hat \psi_\vec{k}$ implies that the quadratic mode energies $\menergy_\vec{k}$ are distributed exponentially with a mode dependent non-equilibrium temperature $T_\vec{k} := \langle \menergy_\vec{k} \rangle_t$  (Fig.~\ref{fig:statistics}a). The Boltzmann constant $k_\text{B}$ is set to unity throughout, defining the units for the temperature, {and all time averages $\langle \cdot \rangle_t$} are calculated after the initial relaxation period. 
The number density function $\ndensity({\varepsilon}) {= \sum\nolimits_\vec{k} \langle \delta (e_\vec{k} - \varepsilon) \rangle_t}$ of modes with energy ${\varepsilon}$  can be expressed as superstatistics \cite{Beck2005} of the individual Boltzmann statistics as  
\begin{equation}\label{eq:mdensity}
\ndensity({\varepsilon}) = \int_{0}^{\infty}\diff \beta\, \mdensity(\beta) \exp(-\beta \varepsilon),
\end{equation}
where $\mdensity(\beta)$ is the number density function of modes with inverse temperature $\beta$.
Since the systems are statistically isotropic, the mode temperatures $T=1/\beta$ can be expressed as a function of the modulus $k$ of the wave vector $\vec{k}$. The localization of the mode energies near the typical pattern wavenumber $k_\text{c}$ allows approximating $T(k)$ as a Gaussian peak with exponentially decaying tails, yielding the following general prediction for the number density function (App.~\ref{sec:radial_distribution})
\begin{equation}\label{eq:superstatistics}
\frac{\ndensity({\varepsilon})}{C_\ndensity} =   \left(
\frac{\exp(-\beta_{\text{t}} {\varepsilon})}{{\varepsilon} \sqrt{\log (\beta_\text{t} T_{\text{g}})}}
+
 \int_{1/T_\text{g}}^{\beta_{\text{t}}} \diff \beta\, 
\frac{ \exp(-\beta {\varepsilon}) }{\sqrt{\log (\beta T_{\text{g}})}}
\right),
\end{equation}
where $C_\ndensity$ is a dimensionless prefactor (see Sec.~\ref{sec:methods} and App.~\ref{sec:radial_distribution} for details of the derivation). The  tail inverse temperature $\beta_{\text{t}}$ is obtained from fitting to the data (see Sec.~\ref{sec:methods} for the fitting procedure), and $T_\text{g} = \max T_\vec{k}$ sets the global temperature scale (Sec.~\ref{sec:methods}). Furthermore, since the systems are statistically translation invariant, the phases are predicted to follow a uniform distribution consistent with the Gaussianity of $\hat \psi_\vec{k}$.

\par
To test these predictions, we measured the {number} density functions of the mode energies {$ \menergy_\vec{k}  $} and the phases $S_\vec{k}= \arg\hat{\psi}_\vec{k} $ in the experiment and simulations. For all three studied systems, we found close agreement between the theory and data~(Fig.~\ref{fig:statistics}): The mode energies of Faraday waves, 2D random scattering and active turbulence follow the universal superstatistics distributions given by {Eq.~\eqref{eq:superstatistics}} with a system specific global temperature $T_\text{g}$, and the phases $S_\vec{k} $ are uniformly distributed in each case. Generally,  {Eq.~\eqref{eq:superstatistics}} can be expected to provide an accurate description whenever the spectral energy is focused within a sufficiently narrow ring (Fig.~\ref{fig:main_figure}c) in Fourier space (Figs.~\ref{fig:breakdown} and \ref{fig:breakdown_quantum} in App.~\ref{sec:breakdown}).


\begin{figure*}
	\includegraphics[width = 2.\columnwidth]{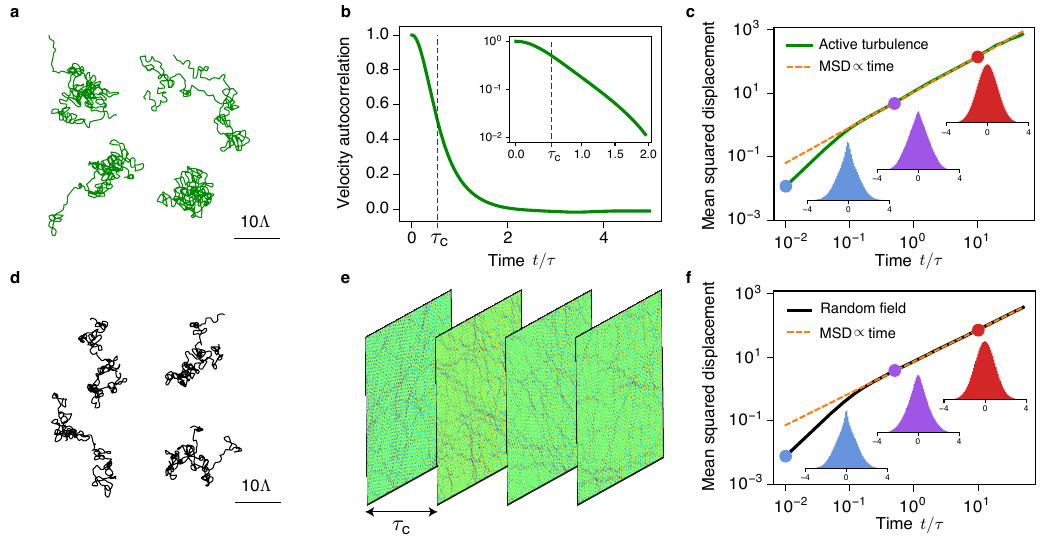}
	\caption{\textbf{Estimating active transport by sampling from monochromatic random fields.} \textbf{a},~Example trajectories of passive tracer particles advected by active turbulent flow solutions (Fig.~\ref{fig:main_figure}a) of the linearly forced Navier-Stokes equations~\eqref{activeNS}. Trajectories are calculated for total time 50$\tau$, where $\tau$ is the typical time scale of pattern growth in Eq.~\eqref{e:tau_GNS}.
	\textbf{b},~Velocity autocorrelations in the solutions of Eqs.~\eqref{activeNS} decay on the order of the pattern growth scale $\tau$ (Sec.~\ref{sec:methods}).
	\textbf{c},~Tracer particles advected by active turbulence move ballistically on time-scales $t \ll \tau$ and diffusively for $t>\tau$.
    \textbf{d}, Sample trajectories of tracer particles in monochromatic random flow fields (Sec.~\ref{sec:methods}) that were periodically updated after time $\tau_c$. 
     \textbf{e}, Vorticity fields $\omega = -\nabla^2 \psi $ corresponding to four stream functions $\psi$  as used in panels (\textbf{d}-\textbf{f}). Stream functions were sampled from superstatistical distributions (Sec.~\ref{sec:methods}) with same {system parameters} as in Movie~4, Fig.~\ref{fig:main_figure}a  and panels (\textbf{a}-\textbf{c}).
	\textbf{f}, Mean squared displacements for tracer particles in monochromatic superstatistics random flow fields agree with those for active turbulence system in panel (\textbf{c}). Panels (\textbf{c}) and (\textbf{f}) show PDFs of the normalized tracer particle displacement at times indicated by the solid circles.
	Mean squared displacement $\langle [\vec{X}_n(t)-\vec{X}_n(0)]^2 \rangle_n $ in (\textbf{c}, \textbf{f}) are based on 100,000 trajectories, respectively. 
	}
	\label{fig:langevin}
\end{figure*}
\par

\section{Application to diffusive transport}
While the emergence of universal statistics in pattern-forming nonequilibrium  systems with scale-selection is rather remarkable in itself, it also opens a path to an efficient field-statistical description of transport processes that avoids explicit simulations of the underlying field dynamics. To demonstrate this, we focus in the remainder on the advection of passive scalars in active turbulence, a process relevant to nutrient transport and mixing in microbial suspensions \cite{2012Sokolov,2012Wensink}. Specifically, we show how one can construct an effective Langevin-type description for tracer dynamics by building on the above results. As a reference process, we simulated the dynamics $\textbf{X}(t)$ of passive particles advected by the solutions $\vec{v}(t,\vec{x})$ of the linearly forced Navier-Stokes equations~\eqref{activeNS}, described by (Fig.~\ref{fig:langevin}a)
\begin{equation}
\label{advection}
    \dot{\vec X}(t)= \vec{v}(t,\vec{X}).
\end{equation} 
Flow field solutions (Movie~4) of Eqs.~\eqref{activeNS} are correlated on short times scales $t<\tau$ and become uncorrelated for $t>\tau$ where $\tau$ is the typical time scale of pattern growth~(Fig.~\ref{fig:langevin}b)  as specified in Eq.~\eqref{e:tau_GNS}. As a consequence, tracer particles advected by solutions  $\vec{v}(t,\vec{x})$ of Eq.~\eqref{activeNS}  move ballistically for $t\ll \tau$ and diffusively for $t> \tau$~(Fig.~\ref{fig:langevin}c). 
\par
To obtain a corresponding Langevin-type description, one has to replace the deterministic Navier-Stokes flow fields  $\vec{v}$ in Eq.~\eqref{advection} by a suitable sequence of randomly generated flow fields. To this end, we constructed stream function fields $\psi(t,\bs x)$  by sampling the Fourier coefficients $ \hat{\psi}_\vec{k} = |\hat{\psi}_\vec{k}| \exp(i S_\vec{k})$. Guided by our above results~(Fig.~\ref{fig:statistics}a), the phases $ S_\vec{k} $ were drawn uniformly from $[0,2\pi)$, and the amplitudes $|\hat{\psi}_\vec{k}|^2$ were sampled from an exponential (Boltzmann) distribution with mean $2 T_\vec{k}/({C_\totalenergy} A |\vec{k}|^2) $, using the mode temperatures $T_\vec{k}$ measured in the active turbulence simulations and the constants {$C_\totalenergy$} and $A$ from  Eq.~\eqref{Fenergy} (Sec.~\ref{sec:methods}). {A new random field with the same spatial correlations is generated at} sampling intervals $ \tau_\text{c} $, which was set to the half-width of the velocity autocorrelation function (Fig.~\ref{fig:langevin}b {and Sec.~\ref{sec:methods}}). The tracer trajectories (Fig.~\ref{fig:langevin}d) obtained by integrating Eq.~\eqref{advection} using the sampled fields (Fig.~\ref{fig:langevin}e) exhibit the same short-term ballistic motion and long-term diffusive behavior (Fig.~\ref{fig:langevin}f) as the reference process  (Fig.~\ref{fig:langevin}c). These results show how transport in pattern-forming systems with scale selection can be approximated through transport in {nearly} monochromatic random fields.

\section{Summary and conclusions}
In conclusion, we found that pattern-forming nonequilibrium systems can be described by emergent superstatistics if the balance of energy injection and dissipation concentrate mode energies near a fixed lengthscale. The above examples illustrate that these conditions are met by a diverse class of quantum and classical systems, regardless of the precise nature of the underlying scale selection mechanisms, driving protocols and mode-mixing nonlinearities. {Furthermore, the three key criteria required by the theory can likely be relaxed leading into further applicability of the theory.}
{Other candidate systems include fluid thermal convection \cite{1993Morris_Spatial}, 
Turing-type chemical reactions \cite{1991Ouyang_transition}, nematic liquid crystals \cite{1989Rehberg}, and
granular materials \cite{1995MeloHexagons,2000HeinrichDoes}.}
The compact statistical representation of dynamical pattern-forming phenomena via monochromatic random fields thus appears to hold promise with regards to the unified description and classification of nonequilibrium dynamics and transport. 

\section{Methods}\label{sec:methods}
\textbf{Faraday wave experiments.}
A schematic of the experimental setup is presented in  Fig.~\ref{fig:SM_schematic}.  
{A circular aluminum bath with inner radius $R=79$ mm}
was filled up to  $H=1.7$ mm with deionized water with density $\rho=998$ kg m$^{-3}$, viscosity $\nu=1$ cSt, and surface tension $\sigma=72.8$ mN m$^{-1}$. A meniscus of characteristic size $l_c\sim2.7$ mm spontaneously formed along the border, where $l_c=\sqrt{\sigma/(\rho g)}$ denotes the capillary length, and $g$ the gravitational acceleration. The bath was mounted on an optical table (Newport SG-34-4 custom breadboard, $3.0'\times 4.0'\times 4.3''$) and vibrated vertically by an electromagnetic shaker (Modal Shop, 2110E) with an external power amplifier (Modal Shop, 2050E09-FS) at acceleration $\Gamma(t)=D\cos(\omega t)$, where  $D$ and $f=\omega/2\pi$ are the prescribed  maximum acceleration and vibrational frequency, respectively. The shaker was connected to the bath by a thin steel rod coupled with a linear air bearing (PI L.P., $4\times4''$ cross section, $6.5''$  long hollow bar) that ensures a spatially uniform vibration to within 0.1$\%$. \cite{Harris2015a} The forcing was monitored through a data acquisition system (NI, USB-6343) with two piezoelectric accelerometers (PCB, 352C65), attached to the base plate on opposite sides of the drive shaft, and a closed-loop feedback  ensured a constant acceleration amplitude to within $\pm0.002$g. \cite{Harris2015a}

At low driving $D$, the fluid remains quiescent. However, as the acceleration is increased beyond a critical driving $D_F$, the so-called Faraday threshold, the layer becomes unstable to a standing field of monochromatic waves \cite{Faraday1831}. The first waves to appear are subharmonic $\omega_F=\omega/2$, with a wavelength $\lambda_F=2\pi/k_F$ prescribed by the water-wave dispersion relation $\omega_F^2=(gk_F+\sigma k_F^3/\rho ) \tanh k_F H$ \cite{Benjamin1954,Douady1990,Bush2015ARFM}.  As the acceleration $D$ is increased beyond $D_F$, nonlinearities excite additional wave modes leading to an order-disorder transition \cite{Tufillaro1989,CrossHohenberg_RMP} beyond which the waves are subject to the chaotic motion discussed in the main text. The experiments were performed at $D=2.500$g and $f=140$\,Hz. At this frequency, the  characteristic wavelength was $\lambda_F=4.60$\,mm and the Faraday threshold $D_F=1.830$g. To investigate the influence of the domain geometry, a range of boundary shapes were tested (see App.~\ref{sec:experiments} and Fig.~\ref{fig:SM_schematic}).

\textbf{Faraday wave field reconstruction.}
The  wave field was measured with a free-surface Schlieren technique \cite{2017Harris,Wildeman2018} that uses a 2D periodic checkerboard pattern as a backdrop to the refractive object of interest (Fig.~\ref{fig:SM_schematic}b). The periodic pattern, which was optically distorted by the presence of the waves, was recorded with a CCD camera (Allied Vision Mako U-130B, $1280\times1024$ pixels) mounted directly above the bath. 
To ensure uniform  normal lighting,  a semi-reflective mirror at $45^{\circ}$ was placed between the CCD camera and the bath, and the bath was illuminated with a diffuse-light LED panel facing the mirror horizontally.
The experiments were recorded in one-minute long videos at 10 frames per second. The relative phase between the frame acquisition and driving signals was adjusted to capture the waves at an instant near maximum average amplitude. The liquid depth $H$ was sufficiently low to prevent phase wrapping \cite{Wildeman2018}. For the spectral analysis presented in the main text we only considered the wave-deformed pattern in a square region of  $93\times 93$ mm centered in the bath (Fig.~\ref{fig:SM_schematic}b). An FFT demodulation method \cite{Wildeman2018} was  then used to reconstruct the 3D surface Faraday waves in that region (Fig.~\ref{fig:SM_schematic}c).
The coefficient $C_\totalenergy$ in Eq.~\eqref{Fenergy} is given by the surface tension  $\sigma$ and the global temperature {with a measured value} $T_\text{g}=4.2$ $ \sigma$cm$^2$.

\textbf{Random scattering simulations.}
We simulated Eq.~(\ref{eq:schrodinger1})  in a doubly-periodic box of size $L = 2 \pi \ell$ using the pseudo-spectral code Dedalus \cite{Burns:2019tf} with $256^2$ Fourier modes.
The initial condition is a uniform plane wave, $\psi(0,\vec{x}) \propto \exp(i \vec{k}_c \cdot \vec{x})$, with wavevector $\vec{k}_c = [64\ell^{-1}, 0]$ and energy $\totalenergy_c = \hbar^2 k_\text{c}^2 / 2 m = \hbar \omega_c$.
The steady potential $V(\vec{x})$ consists of a real Gaussian random field with a Gaussian correlation length of $l_p = 4/ k_\text{c} $.
The potential is rescaled to have zero mean and a standard deviation of $\alpha \omega_c$, where $\alpha$ is adjusted to control the energy concentration in Fourier space. {The parameter $\alpha$ is set to 0.125 for results in Figs.~\ref{fig:main_figure} and \ref{fig:statistics}.}
The simulations are integrated with a second-order implicit-explicit Runge-Kutta scheme using a timestep of $(1/80) \omega_c^{-1}$ until a time of $T = 4 \times 10^5 \omega_c^{-1}$.
For improved numerical conditioning, the system is non-dimensionalized by taking $\hbar = m = \ell = 1$. 
For data in Fig.~\ref{fig:statistics} we let the system relax for time $10^5 \omega_c^{-1}$ before analyzing the data.
The coefficient $C_\totalenergy$ in Eq.~\eqref{Fenergy} is given by $\hbar^2/m$ and {the value for the} global temperature {obtained from the simulations} $T_\text{g}=0.069$ $\hbar^2/L^2/m$. 

\textbf{Active turbulence simulations.}
We simulated Eqs.~\eqref{activeNS} using the pseudo-spectral code described in Ref.~\cite{Slomka2018} using $365^2$ modes. The typical pattern growth timescale is
\begin{equation}
	\label{e:tau_GNS}
	\tau = \left[ 
	\frac{\Gamma_2}{2\Gamma_4} \left( \Gamma_0 - \frac{\Gamma_2^2}{4\Gamma_4} \right)
	\right]^{-1},
\end{equation}
and the width of the energy shell (Fig.~\ref{fig:main_figure}c) reflects the  bandwidth 
$$
\kappa = \left(
\frac{-\Gamma_2}{\Gamma_4} - 2 \sqrt{\frac{\Gamma_0}{\Gamma_4}}
\right)^{1/2}
$$
of active Fourier modes. For all the simulations we set the characteristic vortex size $\Lambda = 1$ (see main text), $\tau = 1$, and $\kappa = 0.3/\Lambda$. The simulations are performed in a periodic box with size $(100\Lambda \times 100\Lambda)$ corresponding to roughly $100^2$ vortices. The time step used for all the simulations was $0.01\tau$. 
The simulations are initialized with a random stream function $\psi$ taking uniformly distributed random values between zero and $10^{-6}$. After this the simulation is run for 10$\tau$ ($10^3$ time steps) to ensure full development of pattern turbulence before any statistical analysis. For the data in \ref{fig:statistics} we ran the simulation for $50,000\tau$ ($5\times 10^6$ time steps) creating 10,000 outputs at uniform time intervals. The panels in Fig.~\ref{fig:main_figure} corresponding to active turbulence simulations are taken from a representative time step after the initial relaxation.  The coefficient $C_\totalenergy$ in Eq.~\eqref{Fenergy} is given by the mass density of the suspension and the global temperature $T_\text{g}=530$ $\rho \Lambda^4/\tau^2$ {obtained from the simulations}.

\textbf{Low-energy cutoff for data in Fig.~\ref{fig:statistics}.} In order to introduce a non-numerical cutoff for small energy modes, we only analyze modes with $T_\vec{k} > 5 \cdot 10^{-3} T_\text{g}$. Note that including all the $k$-modes would result in a large number of nearly zero energy modes whose exact number depends on the real space discretization. Introducing this cutoff allows for turning $\ndensity$ in  Eq.~\eqref{eq:superstatistics} into a probability density function with proper normalization. We point out that similar choices with normalization need to be made with Bose-Einstein number density, which also diverges as $\propto 1/{\varepsilon}$ at small energies.

{
	\textbf{Global temperature $T_\text{g}$.}
	In order to reduce error due to finite time simulations, we calculate the global temperature $T_\text{g}$ from radially averaged mode energies. We define $e_k$ to be the average of energies $e_{\vec{k}'}$ with $||\vec{k}'|-k|<\Delta k/6$, where $\Delta k$ is the discretization size of $k$-modes. We set $T_\text{g} = \max_k \langle e_k \rangle_t$, where the time average is taken after initial relaxation. }

\textbf{Fitting the inverse tail temperature $ \boldsymbol{\beta}_{\text{t}} $.}
Here we describe the fitting procedure for the inverse tail temperature in Eq.~\eqref{eq:superstatistics}. We introduce a splitting of the {number density $ \ndensity $} in terms of the ring and the tail contributions. Let
{
	\begin{equation}
		\ndensity({\varepsilon}) = \ndensity_\text{t}({\varepsilon};\beta_\text{t}) + \ndensity_\text{r}({\varepsilon};\beta_\text{t}),
\end{equation}}
\noindent
where {$ \ndensity_\text{t} $ and $ \ndensity_\text{r} $} correspond to the first and the last term, respectively, in Eq.~\eqref{eq:superstatistics}. 
{The inverse temperature $ \beta_\text{t} $ in our systems is $2-5$ times $1/T_\text{g}$.} 
This implies that {$ \ndensity_\text{t} $} only contributes to the small energy scales of the number density function {i.e. $\ndensity({\varepsilon}) \approx \ndensity_\text{r}({\varepsilon})$ for ${\varepsilon} \gg 1/\beta_\text{t}$}. 
Let {$ \ndensity_\text{e} $} be the empirical number {density function with the low-energy cutoff} obtained from simulations or experiments.  
We start by making a  guess $ \beta_{\text{t}} = \beta_{\text{t}}^{(0)} $ to calculate $ \ndensity_\text{r} $. The {prefactor} {$ C_{\ndensity} $} can be obtained by matching the integrated number densities {$ \int_{{\varepsilon}_0}^{\infty} \diff {\varepsilon} \ndensity_\text{e} \approx \int_{{\varepsilon}_0}^{\infty} \diff {\varepsilon} \ndensity_\text{r} $ with $ {\varepsilon}_0 \gg 1/\beta_\text{t}$.}  
This lets us calculate
\begin{equation}
		\log \left[\left( \ndensity_\text{e}({\varepsilon}) - \ndensity_\text{r}({\varepsilon};\beta_\text{t}^{(0)}) \right)x\right] \approx -\beta_\text{t}^{(1)} {\varepsilon} + \text{constant}
\end{equation}
for ${\varepsilon} \ll 1/\beta_\text{t}$.
Now the parameter $ \beta_\text{t}^{(1)} $ can be obtained from a linear fit for sufficiently small values of $ {\varepsilon} $ and we can start the process again by calculating {$ \ndensity_\text{r} $} using $ \beta_\text{t}^{(1)} $. 
We continue this process until $ \beta_\text{t} $ converges. The data $\text{PDF}({\varepsilon}) \propto \ndensity_\text{e}({\varepsilon})$ shown in Fig.~\ref{fig:statistics}b is normalized such that $\int \diff {\varepsilon} \, \text{PDF}({\varepsilon}) = 1$. This normalization is possible due to the low-energy cutoff as explained in the previous paragraph. The values for $\beta_\text{t}T_\text{g}$ are approximately 5.42, 2.67, and 2.27, for the Faraday wave, random scattering, and active turbulence systems, respectively. 

\textbf{Colorbars in Figure~\ref{fig:main_figure}.} 
The color bar limits in Fig.~\ref{fig:main_figure}a and c are the maxima and the minima of the observed fields. In panel b, the color bars range from  lowest 2.5 percentile up to the 97.5 percentile of the mode energy data, respectively, to increase visibility of the characteristic scars.

\textbf{Tracer particle dynamics.} The active turbulence system was ran for $50\tau$ after initial relaxation creating 4000 time frames. Other simulation details are as described before. We define the velocity autocorrelation as 
$\langle \vec{v}(0,\vec{x})\cdot \vec{v}(t,\vec{x}) \rangle_\vec{x}/ \langle |\vec{v}(0,\vec{x})|^2 \rangle_\vec{x}$ {of the flow field velocities}, where the averages are taken over velocities {at points $\Lambda$ away from each other}. The corresponding random field is created by sampling the squared amplitudes of the Fourier coefficients of the stream function $|\hat{\psi}_\vec{k}|^2$ from a exponential distribution with mean $2 T_\vec{k}/(C_\totalenergy A |\vec{k}|^2)$, while the phases are sampled uniformly. Here the temperatures $T_\vec{k}$ are obtained as energy averages $\langle e_\vec{k} \rangle_t$ from the simulation. For both of the systems, we set up random initial conditions for 100,000 tracer particles to gather the statistics of the mean squared displacement. To update the positions of the particles in time we use a central difference method where the position $\vec{X}_{t_n}$ at discrete time $t_n$ is calculated as $\vec{X}_{t_n} = \vec{X}_{t_{n-1}} + \Delta t (\vec{v}(t_{n-1},\vec{x})+\vec{v}(t_n,\vec{x}))/2$. Here $\Delta t = 0.01 \tau$ is the time step between the time frames and the position $\vec{x}$ corresponds to the finite grid point closest to $\vec{X}_{t_{n-1}}$.


\vspace{0.3cm}\noindent
\textbf{Acknowledgments\\}
We thank  Tapio Ala-Nissil\"a, Geoffrey Vasil and Martin Zwierlein for helpful discussions. This work was supported by a Complex Systems Scholar Award from the James S. McDonnell Foundation (J.D.), the Robert E. Collins Distinguished Scholarship Fund (J.D.), NSF Award DMS-1952706 (J.D.), and Sloan Foundation Grant G-2021-16758 (J.D.). P.J.S. gratefully acknowledges financial support from the NSF (CAREER award CBET-2144180) {and  Alfred P. Sloan Foundation (Sloan Research Fellowships)}. V.H. has been supported by the Academy of Finland via the project Nos. 339228 and 358878.

\vspace{0.3cm}\noindent
\textbf{Author contributions\\}
V.H., J.S. and J.D. developed the theory. V.H. performed analytical calculations, numerical simulations of the GNS model and statistical analysis. K.J.B. contributed the random scattering simulations. A.J.A. and P.J.S. performed the  experiments and Faraday wave reconstruction. V.H. and J.D. wrote the paper with input from all co-authors.

\vspace{0.3cm}\noindent
\textbf{Data availability\\}
All data that support the plots within this paper and other findings of this study are available from the corresponding authors upon reasonable request.



\appendix

\section{Symbols and abbreviations}
%
\begin{tabular}{l|l}
	\hline
	GFF & Gaussian free field \\
	$\psi ({t,}\vec x)$ & Field observable at {time $t$} and position $\vec x$  \\
	$\hat{\psi}_\vk{(t)}$ & Fourier coefficient of $ \psi $ at wave vector $ \vk $ \\ & {and time $t$} \\ 
	$k_\text{c}$ & Characteristic wave number of patterns  \\
	$ \totalenergy {(t)} $ & Total field gradient energy {at time $t$} \\
	$ \menergy_\vk{(t)} $ & Field gradient energy contribution of \\
	&Fourier mode $ \vk $ { at time $t$}  \\
	{$T(k)$} & {Isotropic average mode temperature $\langle e_\vk \rangle$} \\
	{$\beta(k)$} & {Inverse average mode temperature $1/T(k)$} \\
	{$A$} & {Area of the system} \\
	{$\ndensity (\varepsilon)$} & {Number density of $k$-modes with energy $\varepsilon$} \\
	{$\mdensity(\beta)$} & {Number density of $k$-modes with inverse} \\ & {temperature $\beta$} \\
	{$\mu$} & {Volume of a grid point in Fourier space} \\
\end{tabular}
\vspace{2mm}

\noindent
{This list is not intended to be comprehensive. Instead, only the symbols and abbreviations used in multiple sections are included. %
}

\section{Experiments}\label{sec:experiments}
\begin{figure*}[]
	\centering
	\includegraphics[width=1\textwidth]{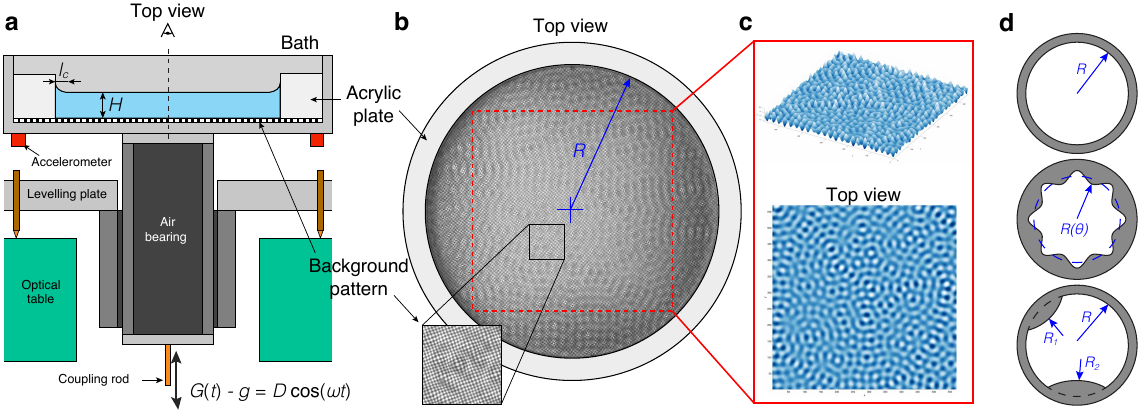}
	\caption{Schematic of the experimental setup. 
		(a) The fluid bath was vibrated with an electromagnetic shaker connected by a thin rod coupled with a linear air bearing. The forcing acceleration was monitored and maintained through two piezoelectric accelerometers and a PID feedback loop.
		(b) Schematic of the test cell where the liquid was confined to a circular bath. A free-surface Schlieren technique was used to reconstruct the interface height by demodulating the optical distortion induced by the Faraday waves in a checkerboard pattern located at the bottom of the bath.
		(c)
		Reconstructed surface height shows weakly chaotic Faraday wave patterns.
		(d) Bath geometries with different rotational symmetries investigated experimentally.
	}
	\label{fig:SM_schematic}
\end{figure*}

Classical and quantum billiards become integrable with certain regular boundaries such as the circular one used in this work preventing thermalization of the system \cite{stockmann_1999}. 
This raises the question of whether the regular boundary might influence the mixing of the energy modes of the Faraday surface waves. 
We thus conducted a systematic study in which we tested a range of bounding shapes of varying degrees of rotational symmetry and size (Fig.~\ref{fig:SM_schematic}d). Specifically, we performed experiments with containers whose inner bounding geometry were (i) circular, (ii) oscillatory, with 8-fold symmetry, and (iii) asymmetric. In all cases, the size of the bath was much larger than the characteristic Faraday wavelength $\lambda_\mathrm{F}=4.6$ mm.

We first conducted experiments with a circular bath of radius $R=79$ mm. The results of these experiments are those discussed in the Main Text. 
We repeated the experiment with a oscillatory boundary  defined by
$R(\theta)=R_b+A_b\cos(8\theta)$ (Fig.~\ref{fig:SM_schematic}b,d). We tested two configurations to also assess the role played by the size of the confining oscillations. In the first one, we considered an oscillatory geometry with a relatively small amplitude given  by $R_b=66.8$ mm and $A_b=8.1$ mm. 
In the second configuration, we increased the size of the bounding oscillations by setting $R_b=60.2$ mm and $A_b=10.6$ mm, and repeated the experiment keeping the rest of the parameters constant. We found that the wave number energy distribution was slightly more isotropic for the smaller oscillations of the boundary layer.

To further test the role of the bath geometry, we conducted experiments in an asymmetric container with two circular protuberances of different size, $R_1$ and $R_2$ (Fig.~\ref{fig:SM_schematic}d). Once again, we tested two configurations: one with relatively smaller defects, given by $R_1=59.3$ mm and $R_2=54.3$ mm, and another one with larger defects, given by $R_1=46.3$ mm and $R_2=39.8$ mm. We note that varying the azimuthal position of the circular protuberances did not have a significant effect on our results.
We observed again that the larger perturbation introduced significant anisotropies in the system, while the effect of the smaller perturbation was more subtle.

We also performed additional tests to ensure that the experimental results were not affected by transient effects or initial conditions. Before each experimental run, we stopped the bath vibration and manually stirred the liquid to make sure that hysteresis did not play a role in our results. We then gradually increased the acceleration and left the bath vibrating for 5 minutes before starting the recording process. We repeated this protocol up to 6 times and confirmed that the emergent statistics results were the same in all cases.

After careful analysis of the circular energy distribution in Fourier space as well as statistical correlations between the modes, we concluded that the circular boundary gave the best results. The experiments with anisotropic perturbed boundaries showed larger autocorrelation times $t_\text{ac}$ (see the following section) for the time evolution of the mode energies as well as persistent anisotropies in the temperatures (Sec.~\ref{sec:radial_distribution}) of the modes. It must be emphasized that, even for the system with a circular boundary, the experiments have to be run sufficiently long to ensure that finite-time correlations do not pollute the data. We thus conclude that the theory works well for the Faraday wave system with a circular boundary as long as the observation times are much larger than the autocorrelation time $t_\text{ac}$ described in the following section. 


\section{Statistical independence and Gaussian probability distribution}\label{sec:correlations}
\begin{figure*}[]
	\centering
	\includegraphics[width=2\columnwidth]{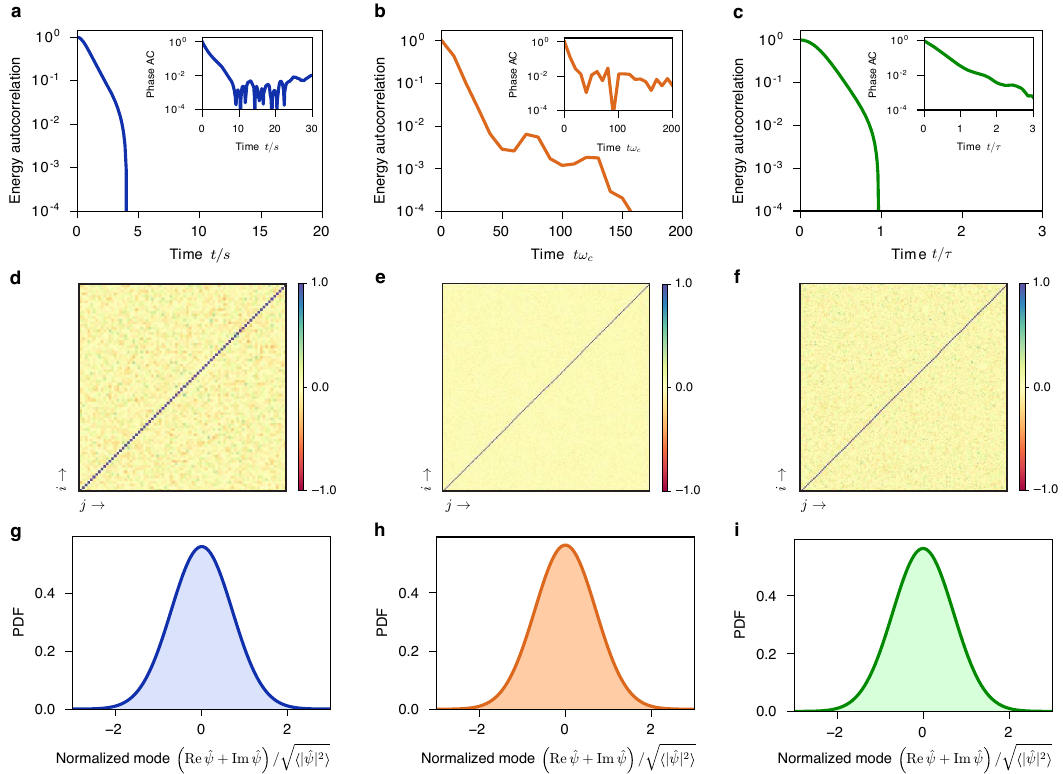}
	\caption{\textbf{a}-\textbf{c}: Autocorrelation functions for the mode energies  $ \menergy_\vk $ on the active ring. The insets show the autocorrelation functions for the  phases $ \mphase_\vk $. \textbf{d}-\textbf{f}: Pair correlations between mode energies $ \menergy_{\vk (\varphi_i)} $ at the critical wave number $ k_\text{c} $. Here $ \varphi_i $ denotes the discrete angle coordinate; a single pixel at $(i,j)$ shows the pair correlation between modes at wave numbers $ \vk = (k_\text{c} \cos \varphi_i, k_\text{c} \sin \varphi_i) $ and $ \vk' = (k_\text{c} \cos \varphi_j, k_\text{c} \sin \varphi_j) $, where $ \varphi_i $ and $ \varphi_j $ are uniformly spaced on the interval $ (-\pi,\pi) $. \textbf{g}-\textbf{i}: The probability density functions of Fourier modes normalized by the variance plotted against a Gaussian fit showing that the modes are normally distributed. We employ a mode cutoff scheme for data in panels \textbf{a}-\textbf{c} and \textbf{g}-\textbf{i} explained in the Methods section of the Main Text. The columns in this Figure show the results for the Faraday wave, the random scattering, and the active turbulence systems, respectively. 
	}
	\label{fig:autocorrelation}
\end{figure*}

The analysis presented in this work is based on the assumption of statistical independence of the mode energies at sufficiently large time scales. Fig.~\ref{fig:autocorrelation}a-c shows the autocorrelation functions 
\begin{equation}\label{eq:autocorrelation_function}
	\text{Aut} (t;\menergy_{\vec{k}}) = \frac{\langle (\menergy_{\vk}(t_0) - \bar{\menergy}_{\vk})(\menergy_{\vk}(t) - \bar{\menergy}_{\vk}) \rangle}{\langle (\menergy_{\vk}(t_0) - \bar{\menergy}_{\vk})^2 \rangle}
\end{equation}
for the energies $ \menergy_\vk $ and the modes $ \mphase_\vk $ for the different systems. 
The reference time $ t_0 $ is chosen to be large enough to avoid capturing any possible transient dynamics occurring after the systems are initialized. 
Here $ \mphase_\vk = \arg \hat{\psi}_\vk $ and $ \bar{\menergy}_\vk = \langle \menergy_\vk \rangle $. 
For the quantum system the autocorrelations decay exponentially with a characteristic time of about $ t_{\text{ac}} = 20 \omega_c^{-1} $ while for the Faraday wave and the active turbulence systems the decay is superexponential with autocorrelation times of $ t_{\text{ac}} = 4$ seconds and $t_{\text{ac}} =  \tau $, respectively. 
Fig.~\ref{fig:autocorrelation}d-e show the pair correlations 
\begin{equation}\label{eq:pair_correlation_function}
	{\text{Cor}(\vk,\vk'; \menergy_\vk)}=
	\frac{\left\langle (\menergy_{\vk}-\bar{\menergy}_\vk) (\menergy_{\vk'} -\bar{\menergy}_{\vk'})\right\rangle_t}{ \sqrt{\langle (\menergy_{\vk}-\bar{\menergy}_\vk)^2\rangle_t \langle (\menergy_{\vk'}-\bar{\menergy}_{\vk'})^2\rangle_t }}
\end{equation}
on the active ring. The averages are taken over time scales larger than $ t_\text{ac}$ showing that the modes are uncorrelated at sufficiently large time scales. 

Figure~\ref{fig:autocorrelation}\textbf{a}-\textbf{c} confirm that  the correlations of $\menergy_\vec{k}$ and $S_\vec{k}$ decay exponentially for the systems studied here. Panels \textbf{d}-\textbf{f} show that the mode energies on the ring are not correlated. Finally, panels \textbf{g}-\textbf{i} give the observational verification for the Gaussianity of the Fourier coefficients of the fields $\psi$ for the different systems. In Sec.~\ref{sec:cumulants} we give a heuristic argument for the Gaussianity of the fields.

\section{Gaussian Free Field (GFF)}
\label{sec:GFF}

\noindent
In the previous section we presented data showing that $ \hat{\psi}_{\vk} $ are approximately normally distributed and independent. This implies that the field $ \psi(\vec x) $ is approximated by a Gaussian Free Field. In this section we will outline some important properties of the GFFs. For a more thorough non-technical primer to GFFs, we refer the reader to Refs.~\cite{hristopulos2020random,monin2007,mccomb2014}.

We consider random fields $ \psi (\vec x): \mathbb{T}^d \to \C^d $ {in a periodic hypercube $\mathbb T ^{d}$} defined as 
\begin{equation}\label{eq:psi_definition}
	\psi (\vec x) = \sum_{\vec k} \hat \psi _{\vec k} e^{i \vec x \cdot \vec k} = \sum_{\vec k} a_{\vec k} (X_{\vec k} + i Y_{\vec k}) e^{i \vec x \cdot \vec k},
\end{equation}
where $ a_{\vec k} $ are {non-negative coefficients} and $ X_{\vec k} $ and $ Y_{\vec k} $ are independent and identically distributed with centered normal distribution $ N(0,1) $. Here $ \vec k = (n_1,n_2,\ldots, n_d) \frac{2\pi}{L} $, where $ L $ is the length of a side of the hypercube {$\mathbb T ^{d}$} (with periodic boundaries) and $ n_i \in \Z$. 

We further assume that the field is statistically translation invariant {(homogeneous)}. This is reflected in the two-point covariance function
\begin{equation}\label{eq:covariance_definition}
	\tilde C(\vec x,\vec y) := \left \langle \psi (\vec x) \psi (\vec y)^* \right\rangle
\end{equation}
as a translation property
\begin{equation}\label{translation_property}
	\tilde C(\vec x + \vec{y}, \vec y) = \tilde C(\vec x,0) =: C(
	\vx)
\end{equation}
for all $ \vec x, \vec y \in \mathbb{T}^d $. 

Next we will show a connection between the coefficients $ a_{\vec k} $ and the Fourier coefficients $ \hat{C}_{\vec k} $ of $ C(\vec x) $. Because of the translation invariant property of the covariance function we have
\begin{equation}\label{f1}
	V C(\vec x) = \left\langle \int_{\T^d} \diff \vec y \psi(\vec x + \vec y) \psi (\vec y)^* \right \rangle,
\end{equation}
where $ V $ is the volume of $ \T^d $. We can write this in terms of the Fourier coefficients $ \hat \psi _{\vec k} $ by using Plancherel theorem as
\begin{equation}
	V C(\vec x) = \left \langle V \sum_{\vec k} e^{i \vec k \cdot \vec x} |\hat \psi _{\vec k}|^2
	\right \rangle.
\end{equation}
It follows that 
\begin{equation}\label{eq:waejfip}
	C(\vec x) = \sum_{\vec k} e^{i \vec k \cdot \vec x} a_{\vec k}^2 \left( 
	\langle X_{\vec k}^2 \rangle + \langle Y_{\vec k}^2 \rangle 
	\right) = \sum_{\vec k} 2 a_{\vec k}^2 e^{i \vec k \cdot \vec x}.
\end{equation}
The last equality holds because $ X_{\vec k}, Y_{\vec k} \sim N(0,1) $. We see that the Fourier coefficients $ \hat{C}_{\vk} $ can be expressed as
\begin{equation}\label{eq:correlation_fourier}
	\hat{C}_{\vec k} = 2 a_{\vec k}^2.
\end{equation}

Another consequence of the translation invariant property is that $ \psi (\vec x) $ are same in distribution for all $ \vec x \in \T^d $. We can calculate the distribution
\begin{equation}\label{eq:groaiwo}
	\begin{split}
		\diff \mathbb{P}(\Re \psi (\vec x) = \xi) &= \diff \mathbb{P}(\Re \psi (0) = \xi) \\
		&= \diff \mathbb{P}\left(\sum_{\vec k} a_{\vec k} X_{\vec k} = \xi \right) \\
		&=: f_{\psi} (\xi) \diff \xi.		
	\end{split}
\end{equation}
Since $ \psi(0) $ can be expressed as a linear combination of independent random variables, the Fourier transform of $ f_{\psi} $  can be written using the probability density function $ f_{X} $ of $ X_{\vec k} $ as 
\begin{equation}\label{eqgwoignwe}
	\hat f _{\psi} (q) = \prod_{\vec k} \hat{f}_X (a_{\vec k} q) = \exp \left(
	-\frac{1}{2} \sum_{\vec k} a_{\vec k}^2 q^2
	\right),
\end{equation}
where the Fourier transform is defined as
\begin{equation}\label{eq:abreoinb}
	\hat{f} (q) = \int_{\R} \diff x \exp(-i q x) f(x).
\end{equation}
We see immediately that the variance is given by
\begin{equation}\label{awgoihwe}
	\operatorname{Var}(\Re \psi (\vec x)) = \sum_{\vec k} a_{\vec k}^2 =  \frac{1}{2} \sum_{\vec k} \hat{C}_{\vec k} = \frac{1}{2} C(0),
\end{equation}
which, by a similar consideration for the imaginary part, implies that $ \langle |\psi(\vec x)|^2\rangle  = C(0) $ as we should expect from the definition of $ C $ given by Eq.~\eqref{eq:covariance_definition} and the translation property Eq.~\eqref{translation_property}.

We note here that
\begin{equation}
	\menergy_\vk \propto \frac{1}{2} k^2  |\hat{\psi}_\vk|^2 
\end{equation}
is exponentially distributed due to the definition of $\hat \psi_\vk$ (Eq.~\eqref{eq:psi_definition})
and
\begin{equation}\label{eq:temperature_correlations}
	\langle \menergy_\vk \rangle \propto \frac{1}{2} k^2 \langle |\hat{\psi}_\vk|^2 \rangle = \frac{1}{2} k^2 \hat{C}_\vk.
\end{equation}
This means that the Fourier coefficients of the covariance function are directly related to the average energy in a mode $ \vk $. In the following section we denote the average energies by a mode specific temperature in analogy with the Boltzmann distribution.

\section{Radial energy distribution}
\label{sec:radial_distribution}

\begin{figure}[b]
	\centering
	\includegraphics[width=\columnwidth]{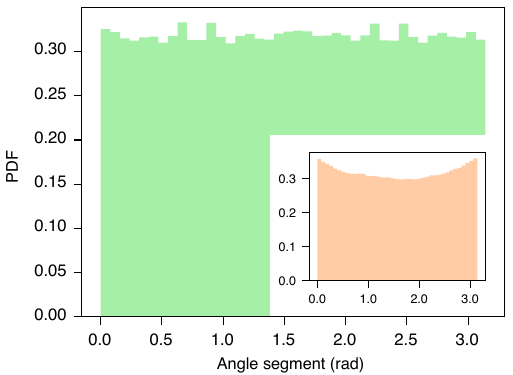}
	\caption{Average mode energies $ \langle \menergy_\vec{k} \rangle $ in an angle segment for the active turbulence system showing isotropic distribution of average energies on the active ring. The inset shows the same data for the random scattering case in which small anisotropy can be noticed at angles 0 and $ \pi $ due to a backscattering effect from the original wave packet at an angle 0 {\cite{Plisson2013}}. The simulation parameters for both of the systems are as described in the Main Text Methods. }
	\label{fig:SM_energy_angle_distribution}
\end{figure}


%
Here we derive the overall energy statistics of the spectral energies $ \menergy_\vec{k} $ using a superstatistical formalism \cite{Beck2003}. The goal is to write the number density of $k$-modes with energy $\varepsilon$ as an integral over inverse temperatures
\begin{equation}\label{eq:superstatistics}
	\ndensity(\varepsilon) = \int_{0}^{\infty} \diff \beta\, \mdensity (\beta) \beta e^{-\beta \varepsilon},
\end{equation}
where $\mdensity (\beta)$ is the number of $k$-modes with inverse temperature $\beta$. This is possible because the mode energies $\menergy_\vk$ are exponentially distributed, as explained in the previous section.


We assume that the long-time averages are isotropic (see Fig.~\ref{fig:SM_energy_angle_distribution}) i.e. they only depend on the modulus of $ \vec{k} $, and define
\begin{equation}\label{eq:radial_energy}
	\langle \menergy_k \rangle _t := \lim\limits_{n \to \infty} \frac{1}{n} \sum_{j=1}^{n} \menergy_{\vec{k}}(t_0 + j\Delta t) = \langle \menergy_k \rangle,
\end{equation}
independent of the starting time $ t_0 $ as long as $ t_0 $ is chosen to be after initial relaxation. Here we assume that the time averages $ \langle \cdot \rangle_t $ are interchangeable with averages $ \langle \cdot \rangle $ over properly randomized initial states due to {ergodicity (latter equality)}.

\begin{figure}[]
	\centering
	\includegraphics[width=\columnwidth]{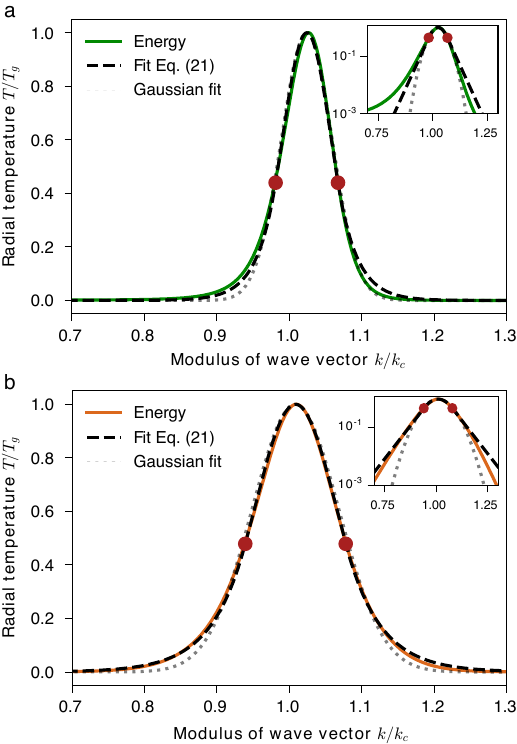}
	\caption{Radial {temperature} $ T/T_{\text{g}} $ at wave number $ k/k_{\text{c}} $ for the active turbulence (\textbf{a}) and the random scattering (\textbf{b}) systems near the activity length scale $ k_\text{c} $. 
		The inset shows the radial temperature on a logarithmic scale revealing the exponential tails. The red dots indicate the points on the curve where the temperature transitions from an exponential to a Gaussian behavior.
		A purely Gaussian fit without the exponential tails is included for reference. 
		The independent parameters $(k_{\text{p}}/k_\text{c},\alpha_{\text{p}}k_\text{c},\beta_{\text{t}}T_\text{g}) $ from this fit are approximately $(1.024,21.0,2.28)$ for the active turbulence case and $(1.008,12.4,2.09)$ for the quantum system. {The deviation from $\kcrit$ of $k_\text{p}$ for the active turbulence case is due to approximations in the definition of $\kcrit$ \cite{Slomka2017b}.}
		The parameters are as described in the Main Text.
	}
	\label{fig:radial_energy distribution}
\end{figure}

The modes $ \menergy_{\vec k} $ are distributed exponentially letting us write
\begin{equation}\label{eq:exponential_average}
	\langle \menergy_k \rangle = \int_{0}^{\infty}\diff \venergyd \frac{\venergyd}{T(k)} \exp \left( -\frac{\venergyd}{T(k)} \right) = T(k).
\end{equation}
$ T(k) $ can be interpreted as a $ k $-dependent local Boltzmann temperature of modes with $ |\vec k| = k $. However, since the systems studied here are not in equilibrium, this temperature might not have all the properties of an equilibrium temperature. 
Fig.~\ref{fig:radial_energy distribution} shows the average energy $ \langle \menergy_k \rangle  = T(k) $ as a function of $ k $ for the active turbulence and the random scattering systems.

Since $\menergy_k$ are exponentially distributed we can write the number density of modes with energy $\varepsilon$ as 
\begin{equation}\label{eq:ndensityk0}
	\begin{split}
		\ndensity(\varepsilon) &=  \sum\nolimits_{\vec k }  \mu \mu^{-1} \beta(k) \exp(-\beta(k) \varepsilon) \\
		&\approx \int_{\mathbb{R}^2} \diff \vec k\, \mu^{-1} \beta(k) \exp(-\beta(k) \varepsilon),
	\end{split}
\end{equation}
where $ \mu = (4 \pi^2)/A $ is the volume of a grid point in Fourier space and $\beta(k) = 1/T(k)$ is the inverse temperature. We assume that the system size $ A $ is sufficiently large to justify the approximate Riemann integral.
Since Eq.~\eqref{eq:ndensityk0} only depends on the modulus of $
\vk$ we have 
\begin{equation}\label{eq:ndensityk}
	\ndensity(\varepsilon)  =  \int_{0}^{\infty} \diff k  \, \frac {2\pi k} {\mu} \beta(k) \exp(-\beta(k) \varepsilon)
\end{equation}
in polar coordinates.

In the following we will introduce a universal approximation for $\beta(k)$ based on the fact that the energy is concentrated near $ k = k_\text{c} $.

Fig.~\ref{fig:radial_energy distribution} shows the radial temperature {T(k)} for the active turbulence and the random scattering cases. The radial temperature distribution can be approximated by a Gaussian peak and two symmetric exponential tails. The Gaussian peak follows from a harmonic expansion of the $ k $ dependent term in the exponential corresponding to a saddle point approximation at the maximum of the narrow peak in the $ k $ dependent temperature distribution.
The exponential tails are justified in Sec.~\ref{sec:exponential_tail}.
We write
\begin{equation}
	T(k) =
	\begin{cases}
		T_\text{t} e^{\alpha_\text{t} (k-k_1)}, & \text{if $ k<k_1 $}; \\
		T_{\text{g}} e^{-\alpha_{\text{p}}^2 (k-k_{\text{p}})^2 }, & \text{if $ k_1 < k < k_2 $}; \\
		T_\text{t} e^{\alpha_\text{t} (k_2 - k )}, & \text{if $ k>k_2 $.}
	\end{cases}
\end{equation}
Symmetry gives $ k_1 + k_2 = 2 k_\text{p} $.
See Fig.~\ref{fig:radial_energy distribution} for a fit to the data for the active turbulence and random scattering cases.
The inverse temperature becomes 
\begin{equation}\label{eq:isotropic_inverse_temperature}
	\beta(k) = 
	\begin{cases}
		\beta_\text{t} e^{-\alpha_\text{t} (k-k_1)}, & \text{if $ k<k_1 $;} \\
		\beta_\text{g} e^{\alpha_{\text{p}}^2 (k-k_{\text{p}})^2 }, & \text{if $ k_1 < k < k_2 $}; \\
		\beta_\text{t} e^{\alpha_\text{t} (k-k_2)}, & \text{if $ k>k_2 $,}
	\end{cases}
\end{equation}
where $ \beta_{\text{t}/\text{g}} = 1/T_{\text{t}/\text{g}} $.


In order to calculate $\ndensity$ we split the domain of integration of Eq.~\eqref{eq:ndensityk} in two intervals, $(0,k_\text{p})$ and $(k_\text{p},\infty)$ and express the integrals in terms of $\beta$. We invert $k(\beta)$ on these intervals resulting in two branches ($k_{-}(\beta)$ and $k_{+}(\beta)$). For $k < k_\text{p}$ we have 
\begin{equation}\label{eq:k_as_beta_0}
	k_{-} = \begin{cases}
		k_\text{p} -  \sqrt{\log(\beta/\beta_\text{g})}/\alpha_{\text{p}},& \text{if $ \beta_{\text{g}} < \beta < \beta_{\text{t}} $;} \\ 
		k_1 - \log(\beta/\beta_\text{t})/\alpha_\text{t}, & \text{if $\beta_{\text{t}} < \beta < \beta(k=0)  $}. \\
	\end{cases}
\end{equation}
For $k > k_\text{p}$ we get
\begin{equation}\label{eq:k_as_beta_1}
	k_{+} = \begin{cases}
		k_\text{p} + \sqrt{\log(\beta/\beta_\text{g})}/\alpha_{\text{p}},& \text{if $ \beta_{\text{g}} < \beta < \beta_{\text{t}} $;} \\
		k_2 + \log(\beta/\beta_\text{t})/\alpha_\text{t}, & \text{if $ \beta > \beta_{\text{t}} $}.
	\end{cases}
\end{equation}
Similarly, 
\begin{equation}\label{eq:k_as_beta_diff}
	\frac{\diff k_{-}}{\diff \beta} = \begin{cases}
		-\frac{1 }{2 \alpha_{\text{p}} \beta \sqrt{\log(\beta/\beta_\text{g})}}, & \text{if $ \beta_{\text{g}} < \beta < \beta_{\text{t}} $;} \\
		-\frac{1}{\alpha_\text{t} \beta}, & \text{if $\beta_{\text{t}} < \beta < \beta(k=0)  $}; 
	\end{cases}
\end{equation}
and
\begin{equation}\label{eq:k_as_beta_diff}
	\frac{\diff k_{+}}{\diff \beta} = \begin{cases}
		\frac{1  }{2 \alpha_{\text{p}} \beta \sqrt{\log(\beta/\beta_\text{g})}}, & \text{if $ \beta_{\text{g}} < \beta < \beta_{\text{t}} $;} \\
		\frac{1}{\alpha_\text{t} \beta}, & \text{if $ \beta > \beta_{\text{t}} $}.
	\end{cases}
\end{equation}
Here $\beta(k=0) = \beta_{\text{t}} e^{\alpha_{\text{t}} k_1}$.

Now we can express the number density as
\begin{equation}
	\frac{\mu}{2\pi} \ndensity(\varepsilon)  = I_1 + I_2 + I_3 + I_4,
\end{equation}
where the integrals on the RHS are
\begin{subequations}
	\begin{align}
		I_1 &= \int_{\beta_{\text{g}}}^{\beta_{\text{t}}} \diff \beta 
		\frac{k_\text{p} -  \sqrt{\log(\beta/\beta_\text{g})}/\alpha_{\text{p}}}{2 \alpha_{\text{p}}  \sqrt{\log(\beta/\beta_\text{g})}} e^{-\beta \varepsilon},  \\
		I_2 &= \int_{\beta_{\text{g}}}^{\beta_{\text{t}}} \diff \beta 
		\frac{k_\text{p} +  \sqrt{\log(\beta/\beta_\text{g})}/\alpha_{\text{p}}}{2 \alpha_{\text{p}} \sqrt{\log(\beta/\beta_\text{g})}} e^{-\beta \varepsilon}, \\
		I_3 &= \int_{\beta_{\text{t}}}^{\beta_{\text{t}} e^{\alpha_{\text{t}}k_1}} \diff \beta  \frac{k_1 - \log(\beta/\beta_\text{t})/\alpha_\text{t}}{\alpha_\text{t}}e^{-\beta \varepsilon}, \\
		I_4 &= \int_{\beta_{\text{t}}}^{\infty} \diff \beta  \frac{k_2 + \log(\beta/\beta_\text{t})/\alpha_\text{t}}{\alpha_\text{t}}e^{-\beta \varepsilon}.
	\end{align}
\end{subequations}

\begin{figure}[]
	\centering
	\includegraphics[width=1.\columnwidth]{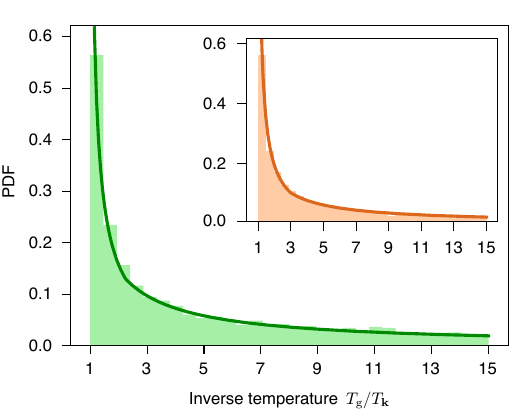}
	\caption{
		Probability density function for the inverse temperature of the active turbulence system {proportional to $\mdensity$}. The solid line shows the theoretical prediction Eq.~\eqref{eq:superstatistics_kernel}. Inset shows the same data for the random scattering case.
		The simulation parameters are as in Main Text and the normalization is explained in the Methods section of the Main Text.
	}
	\label{fig:inv_temp}
\end{figure}

We can safely approximate the upper limit $ \beta(k=0) = \beta_{\text{t}} e^{\alpha_{\text{t}} k_1} \approx \infty $ for these integrals since $\alpha_\text{t} k_1 $ is of the order of $\alpha_{\text{p}} k_\text{c}$ assumed to be large. This approximation will have a negligible contribution to the total number density $\ndensity(\varepsilon)$ as long as $\varepsilon > \beta_{\text{t}}^{-1} e^{-\alpha_{\text{t}} k_1}$ \footnote{The error term produces a divergence $-\log(\varepsilon \beta_\text{t}e^{\alpha_{\text{t}} k_1})/\varepsilon$ when $\varepsilon \to 0$ but since $ e^{-\alpha_{\text{t}} k_1} $ is a very small number, this cannot be verified in simulations.} (see fitted parameters in Fig.~\ref{fig:radial_energy distribution}).
Combining these integrals gives
\begin{equation}\label{eq:ndensity1}
	\begin{split}
		\frac{\mu}{2\pi} \ndensity(\varepsilon) &= \int_{\beta_{\text{g}}}^{\beta_{\text{t}}} \diff \beta \frac{k_\text{p} }{\alpha_{\text{p}} \sqrt{\log(\beta/\beta_\text{g})}} e^{-\beta \varepsilon} \\
		&+ \int_{\beta_{\text{t}}}^{\infty} \diff \beta \frac{k_1 + k_2}{\alpha_{\text{t}}} e^{-\beta \varepsilon},  
	\end{split}
\end{equation} 
where we can use the symmetry $k_1 + k_2 = k_\text{p}$. Writing this in the superstatistical form of Eq.~\eqref{eq:superstatistics} we read
\begin{equation}\label{eq:superstatistics_kernel2}
	\mdensity(\beta)  = \frac{2\pi}{\mu}\begin{cases}
		0, & \beta \in (0,\beta_\text{g});  \\
		\frac{k_\text{p}}{\alpha_\text{p}\beta \sqrt{\log(\beta T_\text{g})}}, & \beta \in (\beta_\text{g},\beta_\text{t}); \\
		\frac{k_1 + k_2}{\alpha_\text{t} \beta} = \frac{2 k_\text{p}}{\alpha_\text{t} \beta}, & \beta > \beta_\text{t}. \\
	\end{cases}
\end{equation}
Requiring continuity for $\mdensity$ implies $ \alpha_{\text{t}} = 2 \alpha_{\text{p}} \sqrt{\log (\beta_\text{t} T_\text{g})} $ resulting in
\begin{equation}\label{eq:superstatistics_kernel}
	\mdensity(\beta) = C_\mdensity \begin{cases}
		0, & \beta \in (0,\beta_\text{g});  \\
		\frac{1}{\beta \sqrt{\log(\beta T_\text{g})} }, & \beta \in (\beta_\text{g},\beta_\text{t}); \\
		\frac{1}{\beta \sqrt{\log (\beta_\text{t} T_\text{g})}}, & \beta > \beta_\text{t}, \\
	\end{cases}
\end{equation}
with the constant $ C_\mdensity = 2\pi k_{\text{p}}/(\alpha_{\text{p}} \mu) = (k_{\text{p}} A)/(2 \pi \alpha_{\text{p}})$. The analytical prediction for $ \mdensity $ is shown against simulation data in Fig.~\ref{fig:inv_temp}.

Using Eqs.~\eqref{eq:superstatistics} and \eqref{eq:superstatistics_kernel} we can evaluate the overall statistics giving 
%
\begin{equation}\label{eq:eowgai}
	\frac{\ndensity(\mathcal{\venergyd})}{C_\mdensity} =  \int_{\beta_\text{g}}^{\beta_{\text{t}}} \diff \beta \frac{ e^{-\beta \venergyd } }{\sqrt{\log(\beta T_\text{g})}}
	+ \int_{\beta_{\text{t}}}^{\infty} \diff \beta \frac{ e^{-\beta \venergyd} }{\sqrt{\log(\beta_\text{t} T_\text{g})}},  
\end{equation} 
which, after calculating the latter integral, becomes
\begin{empheq}[box=\ovalbox]{equation}\label{eq:pdf_large_energy}
	\frac{\ndensity(\venergyd)}{C_\mdensity} =  
	\frac{\exp({-\beta_{\text{t}}\venergyd  })}{\venergyd \sqrt{\log(\beta_\text{t} T_\text{g})}}
	+ \int_{\beta_\text{g}}^{\beta_{\text{t}}} \diff \beta \,
	\frac{ \exp({-\beta \venergyd}) }{\sqrt{\log(\beta T_\text{g})}} .
\end{empheq}
Note that the {prefactor} $C_\ndensity$ used in the Main Text is equal to $C_\mdensity$.

\begin{figure*}[]
	\centering
	\includegraphics[width=100mm]{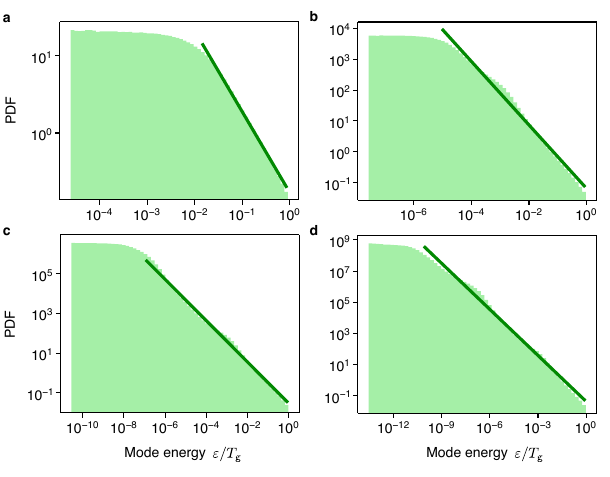}
	\caption{
		Log-log plot of the energy probability density function (proportional to $\ndensity$) for small energies with different energy cutoffs (see Main Text). The tolerances are \textbf{a}: $ 10^{-2} T_\text{g} $, 
		\textbf{b}: $ 10^{-5} T_\text{g} $, \textbf{c}: $ 10^{-8} T_\text{g} $, and \textbf{d}: $ 10^{-11} T_\text{g} $. All linear fits give a power of divergence $ \text{PDF}(\venergyd) \propto \venergyd^{-\gamma} $, 
		where $ \gamma \in [0.98,1.05] $. The plateauing of the distribution near $ \venergyd = 0 $ is due to the cutoff process in the analysis that neglects modes with very low temperatures (energies). 
		The active turbulence simulation parameters are as described in the Main Text Methods. 
	}
	\label{fig:divergence}
\end{figure*}

The second term in Eq.~\eqref{eq:pdf_large_energy} is finite for all $ \venergyd $ implying that the overall asymptotic behavior at $ \venergyd \to 0 $ given by the first term is $ 1/\venergyd $. This result is tested against simulation data in Fig.~\ref{fig:divergence} by fitting an exponential to the energy distribution with small values of $ \venergyd $. 

Note that neither $ \ndensity $ or $ \mdensity $ are integrable over the interval $ (0,\infty) $. This is due to the fact that there are always infinite number of $ k $-modes with zero energy. However, the expectation value of the total energy 
\begin{equation}
	\begin{split}
		&\langle E \rangle = \langle \varepsilon \rangle_{\ndensity}
		= \int_{0}^{\infty} \diff \varepsilon \, \varepsilon \ndensity(\varepsilon) \\
		&= C_\mdensity T_\text{g} \left[ \left( \tilde \beta_\text{t} \sqrt{\log \tilde \beta_\text{t}} \right)^{-1} + \sqrt{\pi} \operatorname{Erf} \left( \sqrt{\log \tilde \beta_\text{t}} \right) \right],
	\end{split}
\end{equation}
where $\tilde \beta_\text{t} = \beta_ \text{t} T_\text{g}$. This is strictly finite
for finite size systems allowing for determining the prefactor $C_\mdensity$. In fact, all the moments $\langle \varepsilon^{n} \rangle_\ndensity$ with $n \geq 1$ are finite implying that Eq.~\eqref{eq:pdf_large_energy} can be used to infer all statistical properties concerning the energy of the system. The small $\varepsilon$ energy statistics are mathematically equivalent to the Thomas-Fermi approximation of a 1-dimensional Bose gas. In photonics it is common to work with the energy spectra $\varepsilon \ndensity(\varepsilon)$ \cite{Weill2017}, which can be normalized to make it a probability density function. 

Note that we do not use the radial temperature data for the results presented in the Main Text. Instead, Eq.~\eqref{eq:pdf_large_energy} is directly fitted against the energy data. {The three independent parameters are, $T_\text{g}$, $C_\mdensity$, and $\beta_\text{t}$. The global temperature is obtained as $T_\text{g} = \max_k T_k$. The constant $C_\mdensity$ can be obtained with any energy (temperature) cutoff as explained in the Methods section of the Main Text.
	The constants $T_\text{g}$ and $C_\mdensity$ set the scale for energies and the number density and can be absorbed into a change of units. The only parameter that changes $\ndensity$ beyond simple scaling is the tail temperature $ \beta_{\text{t}}$. The fitting procedure is detailed in the Methods section of the Main Text. Next we make a few observations about Eq.~\eqref{eq:pdf_large_energy}.}

\subsection{Analysis of $\ndensity$}
Taking $ \alpha_{\text{p}} \to \infty $ while fixing the total energy $ \langle \totalenergy \rangle $ gives a Dirac delta distribution for $ \mdensity $ at some critical $ \beta = \beta_\text{c} $ plus the contribution of infinite number of modes at zero temperature. In this case the total number of modes with energy larger than zero is Boltzmann distributed with inverse temperature $ \beta_\text{c} $. 

Next we will show that the integral term dominating the high-energy behavior of $ \ndensity $ can be effectively approximated for large $ \venergyd/T_\text{g} $. In the following we choose energy units s.t. $T_\text{g}=1$.
We write
\begin{equation}\label{eq:dsgojwjie}
	\int_{1}^{\beta_{\text{t}}} \diff \beta 
	\frac{ e^{-\beta \venergyd} }{\sqrt{\log(\beta)}}
	= \frac{e^{-\venergyd}}{\venergyd} \int_{0}^{(\beta_{\text{t}}-1)\venergyd} \diff \xi\, \frac{e^{-\xi}}{\sqrt{\log(1+\xi/\venergyd)}},
\end{equation}
where $ \xi = \beta \venergyd - \venergyd $. The logarithm $ \log(1 + \xi/\venergyd) \leq \xi/\venergyd $ giving the lower bound
\begin{equation}\label{eq:lower_bound}
	\begin{split}
		\int_{1}^{\beta_{\text{t}}} \diff \beta 
		\frac{ e^{-\beta \venergyd} }{\sqrt{\log(\beta)}} 
		&\geq \frac{e^{-\venergyd}}{\sqrt{\venergyd}} \int_{0}^{(\beta_{\text{t}}-1)\venergyd} \diff \xi\, 
		\frac{e^{-\xi}}{\sqrt{\xi}} \\
		&= \frac{e^{-\venergyd}}{\sqrt{\venergyd}} \sqrt{\pi} \operatorname{Erf}(\sqrt{(\beta_{\text{t}}-1) \venergyd}),
	\end{split}
\end{equation}
where $ \operatorname{Erf} $ is the error function. The error function can be expressed using the complementary error function
\begin{equation}\label{eq:hreaoia}
	\operatorname{Erfc}(\sqrt{x}) = 1 - \operatorname{Erf}(\sqrt{x})
\end{equation}
that has the first order asymptotic formula \cite{abramowitz+stegun}
\begin{equation}\label{eq:gweo}
	\operatorname{Erfc}(\sqrt{x}) = \frac{e^{-x}}{\sqrt{\pi x}} + \BigO\left(\frac{e^{-x}}{x^{3/2}}\right)
\end{equation}
for large $ x $ showing that for large $ (\beta_\text{t}-1)\venergyd $, the error function in Eq.~\eqref{eq:lower_bound} converges exponentially to 1. 

In order to obtain an upper bound we use the Bernoulli inequality
\begin{equation}\label{eq:gori}
	\left(1 + \frac{\xi}{\venergyd}  \right)^{\venergyd} \geq 1 + \xi
\end{equation}
that applies for $ \venergyd \geq 1 $. Taking the logarithm and dividing by the energy gives
\begin{equation}\label{eq:vroeinw}
	\log \left( 1 + \frac{\xi}{\venergyd}   \right) \geq \frac 1 \venergyd \log (1+\xi).
\end{equation}
Using this gives an upper bound 
\begin{equation}\label{eq:upper_bound}
	\int_{1}^{\beta_{\text{t}}} \diff \beta 
	\frac{ e^{-\beta \venergyd} }{\sqrt{\log(\beta)}} \leq \frac{e^{-\venergyd}}{\sqrt{\venergyd}} \int_{0}^{\infty} \diff \xi\, 
	\frac{e^{-\xi}}{\sqrt{\log(1+\xi)}}
\end{equation}
The numerical value of the dimensionless integral on the RHS is about 1.94 whereas $ \sqrt{\pi} \approx 1.77 $. This shows that for large $ \venergyd $ 
\begin{equation}\label{eq:energy_tail_approximation}
	\ndensity(\venergyd) \propto  \int_{1}^{\beta_{\text{t}}} \diff \beta 
	\frac{ e^{-\beta \venergyd} }{\sqrt{\log(\beta)}} \approx \sqrt{\pi} \frac{e^{-\venergyd}}{\sqrt{\venergyd}}.
\end{equation}
Note that because of the units used here the exponential decay rate is set by $ 1/T_\text{g} $.

We make here another remark, namely that in 2 dimensions the real space correlations are long-ranged. Solving from Eq.~\eqref{eq:temperature_correlations} gives
\begin{equation}\label{eq:rgwrio}
	\hat{C}(\vk)  \propto T(\vk)/k^2. 
\end{equation}
An idealized energy concentration on the ring can be modeled by the limit $ \alpha_{\text{p}} \to \infty $  at which $ T(\vk) \propto \delta(k-k_\text{c}) $. We approximate the sums as integrals and calculate the real space correlation function
\begin{equation}\label{eq:real_space_correlations1}
	C(\vec x) \propto \int_{\R^2}\diff \vk \, e^{i \vk \cdot \vec x} \frac{\delta(k-k_\text{c})}{k^2},
\end{equation}
which can be written in polar coordinates as
\begin{equation}\label{eq:real_space_correlations2}
	\begin{split}
		C( \vec x) &\propto \int_0^{2\pi} \diff \theta \int_{0}^{\infty} \diff k\, e^{i k x \cos \theta} k^{-1} \delta(k-k_\text{c}) \\
		&\propto \int_0^{2\pi} \diff \theta e^{i k_\text{c} x \cos \theta}.
	\end{split}
\end{equation}
This integral gives
\begin{equation}\label{eq:real_space_correlations}
	C(\vec x) \propto J_0(k_\text{c} x),
\end{equation}
where $ J_0 $ is the Bessel function of first kind. It has the asymptotic form \cite{abramowitz+stegun}
\begin{equation}\label{eq:bessel_asymptotic}
	J_0 (k_\text{c} x) \approx \sqrt{\frac{2}{\pi k_\text{c} x}} \cos \left(k_\text{c} x -\frac{\pi}{4}\right)
\end{equation}
implying that the correlations decay as $ (k_\text{c} x)^{-1/2} $. This shows that the real space system is thoroughly correlated even if the modes $ \hat{\psi}_{\vec{k}} $ are not.

\pagebreak
\section{Triad interactions} \label{sec:triad_interactions}

\begin{figure}[b]
	\centering
	\includegraphics[width=0.9\columnwidth]{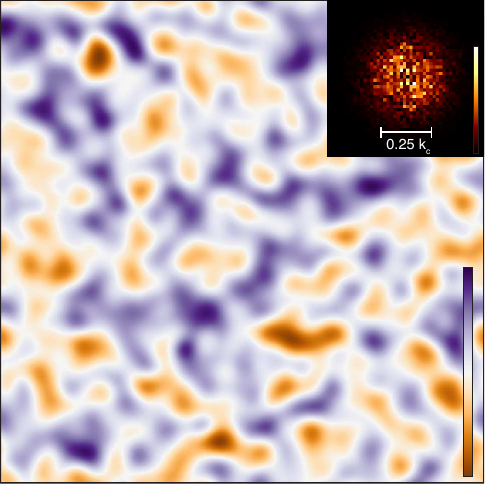}
	\caption{An example of the smooth random potential used for the random scattering simulations described by Eq.~\eqref{eq:schrodinger}. 
		Inset: the modulus of the Fourier transformed potential. {This figure shows the same system as Fig.~2 of Main Text; the potential varies at length scales much larger than the density field $|\psi|^2$.} Color bars are scaled to range from the field minimum to the field maximum. }
	\label{fig:potential}
\end{figure}


Here we introduce \emph{triad interactions} for a more in-depth analysis of our systems in Secs~\ref{sec:cumulants} and \ref{sec:exponential_tail}. Triad interactions appear when a quadratic nonlinearity is expressed in terms of the Fourier coefficients leading into interactions between three modes. For our quantum system this is equivalent to expressing the wave vectors in the momentum basis. 

The triad interactions describe how the Fourier coefficients are dynamically coupled. In the active turbulence case all three modes are Fourier coefficients $ \hat{\psi}_\vk $ whereas in the quantum case interactions between modes $ \hat{\psi}_\vk $ are mediated by the Fourier coefficients $ \hat{V}_\vk $ of the external potential. For more detailed description see e.g.~\cite{Kuhn2005} for the quantum case and \cite{Slomka2018} for active turbulence.


\subsection{Weakly chaotic quantum systems}
\label{sec:weakly_chaotic_quantum}

The quantum dynamics is described  by the Schrödinger equation ($\hbar=1$)
\begin{equation}\label{eq:schrodinger}
	\partial_t \psi(t,\vec{x}) = \frac{i}{2} \Delta \psi(t,\vec{x}) - i V(\vec{x})\psi(t,\vec{x}),
\end{equation}
where $V$ is a smooth random potential (Fig.~\ref{fig:potential}). Writing the time evolution for the Fourier coefficients  
%
\begin{equation}
	\hat{\psi}_\vec{k} =  \frac 1 A \int \diff  \vec{x} \, e^{-i \vec{k} \cdot \vec{x}} \psi(\vec{x}),
\end{equation}
gives the dynamics
\begin{equation}\label{eq:schrodinger_k}
	\partial_t \hat{\psi}_\vk(t) = - \frac{i}{2} k^2 \hat{\psi}_\vk(t) - i (\hat{V} * \hat{\psi})_\vk(t)
\end{equation}
Here $A$ is the are of the system and  $ \hat{V}*\hat{\psi} $ denotes the discrete convolution 
\begin{equation}
	(\hat{V} * \hat \psi)_\vk = \sum\nolimits_{\vq} \hat V_{\vk - \vq} \hat \psi_\vq.
\end{equation}
The triplet $(\vk,\vq,\vk-\vq)$ is called a \emph{triad}.

The potential $V$ is a Gaussian free field (see Sec.~\ref{sec:GFF}) whose Fourier coefficients $\hat V_\vk$ are independent with {zero mean} and variance $\langle |\hat{V}(\vec{k})|^2 \rangle \propto \exp{(-k^2/k_\text{p}^2)} $. Here the inverse correlation length $k_\text{p} = k_\text{c} / 4$, where $ k_\text{c} $ is the wavenumber of the initial state.
The potential is scaled to have variance $\langle V(\vec{x})^2 \rangle = \alpha^2 k_\text{c}^4 / 4$ with $\alpha = 1/8$ for the simulations in the Main Text.


The {kinetic} energy at mode $ \vec{k} $ is given by 
\begin{equation}
	{ \menergy_\vec{k}} =  \frac{1}{2}  A k^2  |\hat{\psi}_{\vec{k}}|^2.
\end{equation}
Multiplying Eq.~\eqref{eq:schrodinger_k} by $ A k^2 \hat{\psi}^*/2 $ and adding the complex conjugate gives the time evolution for the energy of mode $ \vec{k} $
\[ \frac{1}{2} A k^2   { \left( \hat{\psi}^* \partial_t \hat{\psi} + \hat{\psi} \partial_t \hat{\psi}^* \right)}
= \partial_t \left( \tfrac 1 2 A k^2 \hat{\psi}^* \hat{\psi} \right)
= \partial_t {\menergy_\vec{k}}. \]
The right hand side is given by Eq.~\eqref{eq:schrodinger_k}:
\[  k^2 \hat{\psi}^* \partial_t \hat{\psi} = - \frac{i}{2} k^4 |\hat{\psi}|^2 - i k^2 \hat{\psi}^* \hat{V} * \hat{\psi} \]
letting us write 
\begin{equation}\label{eq:energy_evolution}
	\partial_t {\menergy_\vec{k}} =  A k^2 \operatorname{Im}{\left[ \hat{\psi}^* \hat{V} * \hat{\psi} \right]},
\end{equation}
or more explicitly, 
\begin{equation}\label{eq:qenergy_evolution}
	\partial_t {\menergy_\vec{k}} =  A k^2 \sum\nolimits_{\vq} \operatorname{Im}{\left[ \hat{\psi}^*_\vec{k} \hat{V}_{\vk - \vq} \hat{\psi}_{\vq} \right]}.
\end{equation}
Since  $ \hat V $ decays fast as $ |\vec{k} - \vq| $ increases, only modes that are close to each other interact. 

\subsection{Linearly forced Navier-Stokes flow}

For the active turbulence described by Eqs.~(2) of the Main Text, we have in terms of the vorticity $ \omega $ and stream function $ \psi $
\begin{equation}\label{eq:NS_DG}
	\partial_{t} \omega + \nabla \omega \wedge  \nabla \psi  = {\mathcal D} \omega,
\end{equation}
where the $\wedge$-product between two 2D vectors is defined as
\begin{equation}\label{eq:braeoinba}
	\vec{a} \wedge \vec{b} = a_{x} b_{y} - b_{x} a_{y} 
\end{equation}
in terms of Cartesian coordinates so that 
\begin{equation}\label{eq:geaoria}
	\nabla \omega \wedge \nabla \psi = \partial_{x} \omega \partial_{y} \psi - \partial_{x} \psi \partial_{y} \omega, 
\end{equation}
and 
\begin{equation}
	{\mathcal D} = \Gamma_{0}\Delta -\Gamma_{2}\Delta^2 +\Gamma_{4}\Delta^3
\end{equation}
is a differential operator in terms of Laplacians $ \Delta $ {with prescribed coefficients $\Gamma_0$, $\Gamma_2$, and $\Gamma_4$}. 
Vorticity can be obtained from the stream function as $ \omega = -\Delta  \psi $. In Fourier space we have 
\begin{equation}\label{eq:dynamics_Fourier}
	\partial_{t} \hat{\psi}_{\vec{k}}  = \sum_{\vec{k}_1 + \vec{k}_2 = \vec{k}}  (\vec{k}_1 \wedge \vec{k}_2)  \frac{k_1^2}{k^2}  \hat{\psi}_{\vec{k}_1} \hat{\psi}_{\vec{k}_2} + D_\vk \hat{\psi}_{\vec{k}},
\end{equation}
{where
	\begin{equation}\label{eq:dissipation_operator}
		D_\vk = - k^2\left ( \Gamma_0  + \Gamma_2 k^2 + \Gamma_4 k^4 \right )
	\end{equation}
	is the Fourier representation of the operator $\mathcal D$. }
Because the relation $ \vec{k}_1 + \vec{k}_2 = \vec{k} $ is symmetric with respect to changes of indices, we can write the sum as 
\begin{equation}\label{eq:dynamics_Fourier2}
	\partial_{t} \hat{\psi}_{\vec{k}}  = \frac{1}{2}\sum_{\vec{k}_1 + \vec{k}_2 = \vec{k}}  (\vec{k}_1 \wedge \vec{k}_2)  \frac{k_1^2-k_2^2}{k^2} \hat{\psi}_{\vec{k}_1} \hat{\psi}_{\vec{k}_2} + D_\vk \hat{\psi}_{\vec{k}}.
\end{equation}
This shows that modes on the ring $ k_1 = k_2 $ do not contribute to the same triad. 
Writing $ \vec{k}_{1} = \vec{k} - \vq $ and $ \vec{k}_2 = \vq $ gives
\begin{equation}\label{eq:dynamics_Fourier4}
	{\partial_{t} \hat{\psi}_{\vec{k}}  = \sum\nolimits_{\vq} \frac 1 2 (\vec{k} \wedge \vq)  \frac{k^2-2 \vec{k}\cdot \vq}{k^2} \hat{\psi}_{\vq} \hat{\psi}_{\vec{k}-\vq} + D_\vk \hat{\psi}_{\vec{k}}.}
\end{equation}

\par
\begin{figure}[]
	\centering
	\includegraphics[width=0.9\columnwidth]{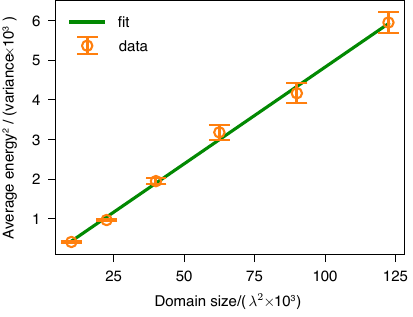}
	\caption{Energy fluctuations as a function of the system size showing that the standard deviation of the total energy $\sqrt{\langle E^2 \rangle_t - \langle E \rangle_t^2 }$ normalized by the average energy $\langle E \rangle_t$ scales as $ 1/\sqrt A $, where $ A $ is the size of the system. The errors are evaluated from finite time sampling by changing the averaging window and choosing the maximum and minimum variances for different window sizes. Simulation parameters are as described in Main Text.}
	\label{fig:energy_deviations}
\end{figure}

The dynamics for the kinetic energy 
\begin{equation}\label{eq:greoaidcf}
	{\menergy_\vec{k}} = \frac{1}{2} A k^2 |\psi_{\vec{k}}| 
\end{equation}
are given by
\begin{equation}\label{eq:NS_Fourier_energy_re}
	\begin{split}
		\partial_t {\menergy_\vk} &= \frac {A} 2 \sum\nolimits_{\vq} ( \vec{k} \wedge \vq ) (k^2-2 \vec{k}\cdot \vq)  \operatorname{Re}{\left[  
			\hat{\psi}_{\vec{k}}^* \hat{\psi}_{\vq} \hat{\psi}_{\vec{k}-\vq}
			\right]} \\
		& + 2 D_\vk {\menergy_\vec{k}}.
	\end{split}
\end{equation}
These dynamics lead to total energy fluctuations that scale out as $ 1/\sqrt A $, where $ A $ is the system size. This scaling is verified by simulations shown in Fig.~\ref{fig:energy_deviations}.

\section{Cumulants of coefficients $\hat \psi_{\vec{k}}$} 
\label{sec:cumulants}
We argue that the fields $\psi(\vx) $ are nearly Gaussian free fields described in Sec.~\ref{sec:GFF}. Showing this rigorously is a remarkably challenging task and shall not be attempted here. In the quantum case there is a proof in the weak coupling limit for Gaussian random potentials at kinetic time scales \cite{Spohn1977} and a more general proof with weak random on-site potentials \cite{Erdos2008}. However, it should be noted that the previous exact results are obtained in the small-coupling limit $\epsilon \to 0$. Therefore for finite $\epsilon$ (potential strength) the proof is only suggestive. 

The argument that we introduce in the following would not work universally for a typical Navier-Stokes equation \cite{mccomb2014} but we argue that the strict length-scale selection mechanism due to the dissipation-activation operator $D$ is the root cause for the Gaussian statistics. 

The argument is based on analysis of cumulants of random coefficients $\hat \psi _\vk$. 
Throughout this section we use averages over randomized initial conditions assuming that these can be interchanged with time averages due to the ergodic property. For detailed properties of cumulants and their time evolution we refer the reader to Ref.~\cite{Lukkarinen2016}. 

The Gaussian distribution is uniquely defined by the property that the \emph{cumulants} $\kappa_n(X)$ of a one-dimensional random variable $X$ disappear for integers $n \geq 3$. Another key property of cumulants is that random variables $(X_1,\ldots,X_k)$ are independent if and only if all the cumulants $\kappa(Y_1,Y_2,\ldots) = 0$, where $Y_j \in (X_1,\ldots,X_k)$ and at least one $Y_j \neq Y_i$. We want to show that these conditions hold approximately for the Fourier coefficients $\hat \psi _\vk$. To this end, we will study the cumulant hierarchies of $\hat \psi_{\vec{k}}$. First, we have to introduce some mathematical notation:

\begin{enumerate}
	\item We denote lists of vectors as $I = (\vk_1,\vk_2, \ldots)$. 
	\item Products of random variables will be denoted as $\hat \psi_{\vec{k}_1}\hat \psi_{\vec{k}_2}\ldots = \hat \psi^I$, where $I$ is defined as before. This lets us write e.g. moments of the form $\langle \hat \psi^{I} \rangle$.
	\item Lists of random variables are written as $\hat \psi_{I} = (\hat \psi_{\vk_1},\hat \psi_{\vk_2},\ldots)$. This is used in particular with cumulants of several variables e.g. $\kappa(\hat \psi_{I}) = \kappa(\hat \psi_{\vk_1},\hat \psi_{\vk_2},\ldots)$.
	\item The cumulant $\kappa_{n}(X)$ is a shorthand for $\kappa(X,X,\ldots,X)$, where $X$ is repeated $n$ times. 
\end{enumerate}

Cumulants for several random variables can be calculated using the moments to cumulants formula
\begin{equation}\label{eq:moments2cumulants}
	\langle \hat \psi^{I} \rangle = \sum_{\pi \in \mathcal{P}(I)} \prod_{A \in \pi} \kappa(\hat \psi_{A}),
\end{equation}
where the sum goes over all the partitions of the list $I$ and the product multiplies different clusters in the partition. Eq.~\eqref{eq:moments2cumulants} can be inverted giving 
\begin{equation}\label{eq:cumulants2moments}
	\kappa ( \hat \psi_I ) = \sum_{\pi \in \mathcal{P}(I)} (|\pi|-1)! (-1)^{|\pi| - 1} \prod_{A \in \pi} \langle \hat \psi^{A} \rangle,
\end{equation}
to define any cumulants in terms of the moments. Again the sum goes over all partitions of $I$ and $|\pi|$ is the number clusters in the partition $\pi$. 
Alternatively, cumulants can be defined using the cumulant generating function 
\begin{equation}
	g_\text{c}(\eta_I;\hat \psi_I) := \log \left( \left\langle \exp\left(\sum_{\vk \in I} \eta_{\vk} \hat \psi_{\vk} \right) \right\rangle \right)
\end{equation}
if all the moments exist. It should be understood here that $g_c$ is not a function of $\hat \psi_I$ in the traditional sense. Instead, it is a function of the underlying joint probability measure of $\hat \psi_I$. The cumulants can be then calculated as
\begin{equation}
	\kappa(\hat \psi_I ) = (\partial_{\eta_{I}} g_\text{c}(\eta_I; \hat \psi_I) )_{\eta_{I} = 0},
\end{equation}
where $\partial_{\eta_{I}} = \partial_{\eta_{\vk_1}} \partial_{\eta_{\vk_2}} \ldots $ with $\eta_{\vk_j} \in I$. Note that the same vectors can be repeated in $I$. 

In order for our field $\psi $ to be near Gaussian, we require that cumulants of the form $\kappa_{n,n}(\hat \psi_\vk, \hat \psi_\vk^*)$ with $n \geq 2$ and $\kappa(\hat \psi_{I})$, where $I$ contains any two distinct vectors, are small.

Later on we will use a translation invariant property of our fields, namely that the $n$-point covariance function
\begin{equation}
	\langle \psi(\vx ) \psi(\vx + \vec{d}_1) \ldots \psi(\vx + \vec{d}_{n}) \rangle = C_{n+1}(\vec{d}_1,\ldots, \vec{d}_{n})
\end{equation}
is independent of $\vx$. This implies a condition for the coefficients $\hat \psi_\vk$:
\begin{equation}\label{eq:resonance_condition}
	\langle \hat{\psi}_{\vk_1}^{(s_1)} \ldots \hat{\psi}_{\vk_n}^{(s_n)} \rangle = 0, \; \text{if} \; \sum_{j=1}^{n} s_j \vk_j  \neq 0,
\end{equation}
where
\begin{equation}
	\hat{\psi}_{\vk_j}^{(s_j)} = \begin{cases}
		\hat{\psi}_{\vk_j} & \text{if $s_j =1$};\\
		\hat{\psi}_{\vk_j}^* & \text{if $s_j =-1$},
	\end{cases}
\end{equation}
i.e. the sign in the resonance condition is flipped for the fields that have been complex conjugated. Two very useful corollaries are  
\begin{equation}
	\langle \hat \psi_\vk \rangle = 0,
\end{equation} 
when $\vk \neq 0$ (in our case also $\langle \hat \psi_0 \rangle = 0$) and 
\begin{equation}
	\langle \hat \psi_{\vk}^* \hat \psi_{\vq} \rangle = 0, 
	\label{eq:resonance_2nd}
\end{equation}
if $\vk \neq \vq$. The same resonance condition holds for cumulants. This can be seen by looking at Eq.~\eqref{eq:cumulants2moments}:
If $\sum_{\vk \in I} s_\vk \vk \neq 0 $, it follows that for each term in the sum in Eq.~\eqref{eq:cumulants2moments}, at least one $A \in \pi$ has the property $\sum_{\vk \in A} s_\vk \vk \neq 0 $. Conversely, if $\sum_{\vk \in I} s_\vk \vk = 0 $, the sum picks up exactly the partitions $\pi$ for which $\sum_{\vk \in A} s_\vk \vk = 0$ for all $A \in \pi$, which is true always for at least $\pi = \lbrace I \rbrace$.

Next we will look at the time evolution of the cumulants. 
Let us assume that the dynamics of the Fourier coefficients of a real field $\psi$ can be written as 
\begin{equation}\label{eq:time_evolution_general}
	\partial_t \hat \psi_{\vk} = D_{\vk} \hat{\psi}_{\vk} + \sum\nolimits _{\vq} K_{\vk \vq} \hat \psi_{\vq} \hat \psi_{\vk - \vq}
\end{equation}
with some linear operator $D_\vk$ and an interaction kernel $K_{\vk\vq}$. 
The dynamics for the cumulants can be written as 
\begin{equation}
	\partial_t \kappa(\hat \psi_{I}) = \sum_{\vk \in I} \langle \partial_t \hat \psi_\vk : \hat \psi^{I \setminus \vk}: \rangle,
\end{equation}
where $: \hat \psi^{I \setminus \vk}:$ is the \emph{Wick polynomial} of the coefficients in the list $I$ with $\vk$ removed. Here it suffices to know that we can expand this moment in cumulants using Eq.~\eqref{eq:moments2cumulants} and the Wick polynomial will have all the partitions with clusters internal to $\hat \psi^{I \setminus \vk}$ removed \cite{Lukkarinen2016}. Plugging in Eq.~\eqref{eq:time_evolution_general} gives 
\begin{equation}
	\begin{split}
		\partial_t \kappa (\hat \psi_{I} ) &= \sum_{\vk \in I} D_\vk \langle \hat \psi_{\vk} : \hat \psi^{I \setminus \vk}: \rangle  \\
		&+ \sum_{\vk \in I} \sum\nolimits_{\vq} K_{\vk\vq} \langle \hat \psi_{\vq} \hat \psi_{\vk - \vq} : \hat \psi^{I \setminus \vk}: \rangle.
	\end{split}
\end{equation}
For the first term the only partition with no internal clusters of $\hat \psi^{I \setminus \vk}$ is the whole cumulant $\kappa (\hat \psi_I )$. The second term gives the whole cumulant plus the coefficients $\hat \psi_{I \setminus \vk}$ split between $\hat \psi_\vq$ and $\hat \psi_{\vk - \vq}$. Thus the whole time evolution can be written as 
\begin{equation}\label{eq:cumulant_time_evolution}
	\begin{split}
		\partial_t \kappa (\hat \psi_{I}) &= D_I \kappa (\hat \psi_{I}) \\
		&+ \sum_{\vk \in I} \sum\nolimits_{\vq} K_{\vk\vq} \left[ \kappa(\hat \psi_\vq, \hat \psi_{\vk - \vq}, \hat \psi_{I \setminus \vk}) \right. \\
		&+ \sum_{J \subset I \setminus \vk} \left. \kappa(\hat \psi_{\vq}, \hat \psi_{J}) \kappa(\hat \psi_{\vk - \vq}, \hat \psi_{J^c}) \right].
	\end{split}
\end{equation}
Here the last sum includes all the sublists $J$ and its complement $J^{c}$ of the list $I \setminus \vk$, and  
\begin{equation}
	D_I = \sum_{\vk \in I} D_\vk.
\end{equation}
The sublists $J$ do not include the empty list or the whole list because due to homogeneity of the field, $\kappa(\hat \psi_{\vq}) = 0$. 
Note that the time evolution of the cumulant is linked to a cumulant that is one degree higher plus a contribution of products of lower order cumulants. The upward expansion term $ \kappa(\hat \psi_\vq, \hat \psi_{\vk - \vq}, \hat \psi_{I \setminus \vk}) $ is always resonant (if $\kappa(\hat \psi_I)$ is resonant) and all the terms in the sum must be accounted for. However, the downward expansion term $ \kappa(\hat \psi_{\vq}, \hat \psi_{J}) $ is resonant for exactly one $\vq$ with the property 
\begin{equation}
	\vq +\sum_{\vec p \in J} \vec{p} = 0.
	\label{eq:sub-resonance}
\end{equation}

In the following we will show that for relevant interactions, the kernel $K_{\vk\vq}$ is small. This corresponds to a weak coupling between the modes and allows for the truncation of the cumulant hierarchy in Eq.~\eqref{eq:cumulant_time_evolution}. 

\subsection{Active turbulence} 
We present the theory first for the active turbulence case, which turns out to be conceptually simpler. 
In this case we have the coefficients $D_\vk$ of the dissipation-activation operator and 
\begin{equation}\label{eq:A_operator}
	K_{\vk\vq} = \frac{1}{2} \vk \wedge \vq \frac{|\vk - \vq|^2 - q^2}{k^2},
\end{equation}
which has the symmetries $K_{\vk\vq} = K_{-\vk,-\vq}$ and $K_{\vk\vq} 
= K_{\vk,\vk - \vq}$ (see Eq.~\eqref{eq:dynamics_Fourier4}).

\begin{figure}
	\centering
	\includegraphics[width=\columnwidth]{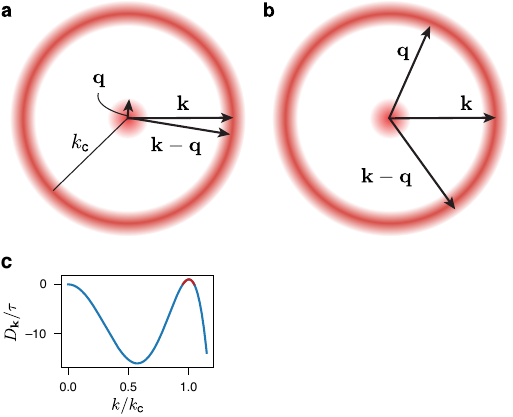}
	\caption{ {
			Dominant interactions consist of triads for which $D_{\vq} + D_{\vk - \vq}$ is positive. These come in two variants illustrated here for mode $\psi_\vk$ on the active ring:
			\textbf{a}, One mode near the origin and another one on the active ring.
			\textbf{b}, Two modes on the active ring. Note that the triad $(\vk, \vq, \vk -\vq)$ sums up to 0 forming a triangle on the plane. 
			Panel \textbf{c} shows the isotropic dissipation-activation  operator $D_{\vk}$ Eq.~\eqref{eq:dissipation_operator} as a function of $k$ with $\kappa/\kcrit = 0.3 / \pi$, the same value used for the system analyzed in the Main Text. }
	}
	\label{fig:triad_interaction}
\end{figure}

We assume that Eq.~\eqref{eq:cumulant_time_evolution} has a stationary solution for $\kappa(\hat \psi_I) $ given by 
\begin{equation}\label{eq:cumulant_time_stationary}
	\begin{split}
		D_I \kappa (\hat \psi_{I}) &= -\sum_{\vk \in I} \sum\nolimits_{\vq} K_{\vk\vq} \left[ \kappa(\hat \psi_\vq, \hat \psi_{\vk - \vq}, \hat \psi_{I \setminus \vk}) \right. \\
		&+ \sum_{J \subset I \setminus \vk} \left. \kappa(\hat \psi_{\vq}, \hat \psi_{J}) \kappa(\hat \psi_{\vk - \vq}, \hat \psi_{J^c}) \right].
	\end{split}
\end{equation}
Eq.~\eqref{eq:cumulant_time_evolution} shows that there is an exponential growth factor if $D_I > 0$. On the other hand if $D_I \ll 0$, there is fast exponential decay. $D_{\vk}$ decreases rapidly away from the origin or the active ring (Fig.~\ref{fig:triad_interaction}) suggesting that \emph{any} cumulants with modes away from the origin or the active ring have to be small. This is an argument for the scale-selection i.e. that all the relevant modes are on the active ring or near the origin.

Looking at modes $\hat \psi_\vq$ and $\hat \psi_{\vk - \vq}$ interacting with $\hat \psi_{\vk}$, where $k$ is near $k_\text{c}$ we have two possible dominant interactions, illustrated in the panels \textbf{a} and \textbf{b} of Fig.~\ref{fig:triad_interaction}:
\begin{enumerate}
	\item[\textbf{a},] One of the interacting modes is near the origin and another one near $\hat \psi_\vk$.
	\item[\textbf{b},] Both $\hat \psi_\vq$ and $\hat \psi_{\vk - \vq}$ are near the active ring.
\end{enumerate}
In both of the cases, the interaction kernel $K_{\vk \vq}$ is small:
\begin{enumerate}
	\item[\textbf{a},] $|\vk \wedge \vq|/k_\text{c}^2 = |\vk \wedge (\vk - \vq)|/k_\text{c}^2 = \mathcal O(q/\kcrit)$ is small. We choose here $\vq$ to be the small mode (see the symmetries of $K$) i.e. $q \ll k_\text{c}$. 
	\item[\textbf{b},] $|\vk - \vq| \approx q \approx k_\text{c}$ implying that $(|\vk - \vq|^2 - q^2)/k_\text{c}^2$ is small. Note that modes with $|\vk - \vq| = q$ i.e. at the same length scale do not interact. 
\end{enumerate}
In conclusion, $K/k_\text{c}^2$ is small for the dominant interactions. 

Another equally important feature of the interactions is that due to the length-scale selection, the sum over $\vq$ in Eq.~\eqref{eq:cumulant_time_stationary} does not contain many significantly large modes. The average number of the squared field $\psi(\vx)$ can be calculated as 
\begin{equation}
	\label{eq:number-of-modes}
	\frac{1}{A} \int \diff \vx\, \langle \psi(\vx)^2 \rangle = \langle \psi(\vx)^2 \rangle = \sum\nolimits_{\vk} \langle |\hat \psi_\vk|^2 \rangle,
\end{equation}
which is independent of the system size. Our simulations show that the average $\sqrt{\langle \psi(\vx)^2 \rangle} \approx 24 \tau^{-1} \kcrit^{-2}$ with the parameters used in the Main Text. This value is obtained if \emph{all} the modes are included. We argue that the fraction of the significantly large modes included in one of the dominant interactions is small. To this end, it is instructive to look at Fig.~\ref{fig:triad_interaction}: For type \textbf{a} interactions, only the triads with $\vq$ in a small neighborhood of the origin will contribute whereas for type \textbf{b} interactions the number of relevant modes is constrained on a small angular segment of the active ring.
This implies that the sum over $\vq$ in Eq.~\eqref{eq:cumulant_time_stationary} contains only a small fraction of large modes contributing to the average $\sqrt{\langle \psi(\vx)^2 \rangle}$.


Next we use Eq.~\eqref{eq:cumulant_time_stationary} to hierarchically solve for the equilibrium value of the cumulant. We conclude the hierarchical expansion in branches where all the remaining cumulants are of the order 2 giving products of the cumulants $\kappa(\hat \psi_{-\vk}, \hat \psi_{\vk}) = \langle |\hat \psi_\vk |^2 \rangle$ with some $\vk$.

As an example, few of the first cumulants have the following hierarchies:
\begin{equation}
	\begin{split}
		(3) &\to (4) + (2)(2) \\
		(4) &\to (5) + (3)(2) \to (2)(2)(2) + \text{h.o.} \\
		(5) &\to (6) + (2)(4) + (3)(3)  \\
		&\to (2)(2)(2)(2) + \text{h.o.}
	\end{split}
\end{equation}
The first row means that expanding a 3rd order cumulant will give 4th order cumulants plus products of 2nd order cumulants. Here h.o. denotes \emph{higher order} terms in products of $(2)$. 

We will first consider the downward expansion terms, for which the maximum order of the cumulants in the subsequent product decreases e.g. $(3) \to (2)(2)$. Let $\bar K$ denote the order of the operator $K$. Per the previous discussion we assume that the value of $\bar K / \kcrit^2$ is small. Now, with every downward expansion we introduce a term $\bar K$. It should be also remembered that due to the resonance condition, only one term in the sum over $\vq$ is retained. Since the maximum degree of the cumulant in a direct downward chain decreases by one with every step, the order of the product of such a chain is $\bar K^{n-2}$. As an example, a downward expansion of a 3rd order cumulant is of the order of $\bar K \langle |\hat \psi_{\vk}|^2 \rangle^2 $ (see Sec.~\ref{sec:exponential_tail} for the calculation). 

Next we will consider the upward expansion. We note here that cumulants are multi-linear satisfying 
\begin{equation}
	\kappa(\alpha \hat \psi_\vq, \hat \psi_J ) = \alpha \kappa(\hat \psi_\vq, \hat \psi_J ) 
\end{equation}
for any real number $\alpha$, vector $\vq$, and list $J$. Since the number of significant modes included in the sum over $\vq$ is not large and the coupling parameter $K$ is small, we argue that the upward expansion of a cumulant of order $n$ scales as
\begin{equation}
	\label{eq:upscaling}
		\sum\nolimits_{\vq} \frac{K_{\vk \vq}}{\kcrit^2} \kappa(\hat \psi_\vq, \hat \psi_{\vk - \vq}, \hat \psi_{I \setminus \vk} ) 
		\sim \bar C \kappa(\hat \psi_{I}),
\end{equation}
where $\sim$ denotes the order of the terms and $\bar C$ is a small \emph{effective} coupling constant. Here we also need the fact that the values of the coefficients $\hat \psi_\vk$ are constrained by the average $\sqrt{\langle \psi(\vx)^2 \rangle} $. Eq.~\eqref{eq:upscaling} ensures that with every upward expansion the cumulant is multiplied by $\bar C$ effectively truncating the upward chains. The conclusion is that the dominant terms will come from the direct downward chains. 


To summarize, we have shown evidence arguing that the upward expansion of the cumulants can be truncated. Moreover, we have shown that the remaining direct downward expansions give the scaling $\bar K^{n-2}$ for cumulants of order $n$ explaining why the high order cumulants are small. 
This supports the observation in Sec.~\ref{sec:correlations} that $\hat \psi_\vk$ is nearly Gaussian. We note that the underlying reason for this behavior is that $D_\vk$ concentrates the modes on the active ring forcing weak coupling between the modes. In this discussion we have omitted the dissipation-activation term $D_I$, which can be equal to zero. Our preliminary investigations suggest that the values of $\vq$, where $D_I$ are small are not important for the dynamics (see Sec.~\ref{sec:exponential_tail}). In order to avoid a lengthy discussion on singular integrals, we have decided to omit this subject from this exposition. 

In classical turbulence problems this method is unlikely to work for the following reasons:
\begin{enumerate}
	\item Although the interaction kernel $K$ is the same, there is no reason for it to be small.
	\item The number of the modes in the sum over $\vq$ can be arbitrarily large without the limitation set by the length-scale selection mechanism.
	\item In high Reynolds number turbulence the average value $\langle \psi (\vx)^2 \rangle $ can be large. 
\end{enumerate}
The last point might be true even with length-scale selection with large enough driving on the active ring compared to dissipation away from the ring (see the widest ring in Fig.~\ref{fig:breakdown}).

\subsection{Random scattering}
\label{sec:cumulant-quantum}
In the quantum case we make a change of variables to
\[ \hat \phi_\vk = \hat \psi_{\vk} e^{i k^2 t/2} \]
corresponding to the interaction picture of quantum mechanics. 
In these variables the time evolution of the modes is determined by
\[ \partial_t \hat \phi_{\vk} = -i \sum\nolimits _{\vq} \hat V_{\vk - \vq} \hat \phi_{\vq} e^{i \frac{t}{2} (k^2-q^2)} \]
and its complex conjugate counterpart. We use the shorthand notation
\begin{equation}
	\hat \phi_{\vk^*} := \hat \phi_\vk ^*.
\end{equation}
Since this equation does not fit in the form given by Eq.~\eqref{eq:time_evolution_general}, we have to modify the time evolution terms for the cumulants. Now we have to analyze cumulants of the form
\begin{equation}
	\begin{split}
		&\partial_t \kappa (\hat \phi_{I}, \hat V_J ) =
		\sum_{\vk \in I} \langle \partial_t \hat \phi_\vk : \hat V^J \hat \phi^{I \setminus \vk} : \rangle \\
		&=
		\sum_{\vk \in I} \sum\nolimits_{\vq}  K_{\vk\vq}(t)^{(s_\vk)} \langle \hat \phi_{\vq}^{(s_\vk)} \hat V_{\vk - \vq}^{(s_\vk)} : \hat V^J \hat \phi^{I \setminus \vk} : \rangle,
	\end{split}
	\label{eq:quantum-cumulant1}
\end{equation}
where, as before, $f^{(s_\vk)}$ is complex conjugated only if $\phi_\vk$ is complex conjugated.
Here 
\begin{equation}
	K_{\vk\vq}(t) = -i e^{i \frac{t}{2} (k^2 - q^2)}
\end{equation}
with the symmetries
\begin{equation}\label{eq:quantum-kernel-symmetries}
	\begin{split}
		K_{-\vk,-\vq}(t) &= K_{\vk\vq}(t), \\
		K_{\vq\vk}(t) &= -K_{\vk\vq}(t)^* = K_{\vk\vq}(-t).
	\end{split}
\end{equation}
Eq.~\eqref{eq:quantum-cumulant1} becomes 
\begin{equation}
	\begin{split}
		&\partial_t \kappa (\hat \phi_{I}, \hat V_J ) \\&=
		\sum_{\vk \in I} \sum\nolimits_{\vq} K_{\vk\vq}(t)^{(s_\vk)} \left[ \kappa(\hat \phi_\vq^{(s_\vk)}, \hat V_{\vk - \vq}^{(s_\vk)},\hat \phi_{I \setminus \vk},\hat V_J)\right. \\
		& + \sum_{\substack{K_1 \subset I\setminus \vk, \\ K_2 \subset J }} \left.\kappa(\hat \phi_\vq^{(s_\vk)},\hat \phi_{K_1}, \hat V_{K_2}) \kappa(\hat V_{\vk - \vq}^{(s_\vk)},\hat \phi_{K_1^c}, \hat V_{K_2^c}) \right].
	\end{split}
	\label{eq:quantum-cumulant2}
\end{equation}
The potential $V$ is a random Gaussian field with long-range Gaussian correlations. This means that $\hat V_\vk$ are independent and normally distributed with zero mean. The field is fully determined by the covariance relation
\begin{equation}\label{eq:potential-covariance}
	\langle \hat V_\vk \hat V_{\vq} \rangle = \delta (\vk + \vq) G_\vk,
\end{equation} 
where the covariance function $G$ is given by
\begin{equation}
	G_\vk = C_G \exp \left (-\tfrac 1 2 \tfrac{k^2}{k_\text{p}^2} \right )
\end{equation}
and the coefficient $C_G$ is determined by
\begin{equation}
	\sum\nolimits _\vk G_\vk = \langle V(x)^2 \rangle = \epsilon^2.
\end{equation}
We define the inverse correlation length $k_\text{p} \ll k_\text{c}$ and the energy scale $\epsilon \ll \tfrac 1 2 k_\text{c}^2$. Both the fields $V$ and $\psi$ are statistically translation invariant implying that the resonance condition \eqref{eq:resonance_condition} holds. In addition, cumulants of the form $\kappa(\hat V_J, \hat \phi_I) = 0$ if the list $(\hat \phi_I)$ does not have an equal number of fields and their complex conjugates. This follows from the gauge invariance of the field $\phi$ with respect to constant phase transformations $\phi \to \phi \exp(i \theta)$. For example, the cumulant $\kappa(\hat V_\vk, \hat \phi_{\vk}^*) = 0$ even as the resonance condition \eqref{eq:resonance_condition} holds.  

Next we integrate Eq.~\eqref{eq:quantum-cumulant2} from $0$ to $t$. Note that there are no correlations of $(\hat \phi_I,\hat V_J)$ at time $t=0$. This gives 
\begin{equation}
	\label{eq:quantum-cumulant-duhamel}
	\begin{split} 
		&\kappa (\hat \phi_{I}, \hat V_J ) \\&= 
		\sum_{\vk \in I} \sum\nolimits_{\vq} \int_{0}^{t} \diff t_1 K_{\vk, \vq}(t_1)^{(s_\vk)} \left[ \kappa(\hat \phi_\vq^{(s_\vk)}, \hat V_{\vk - \vq}^{(s_\vk)},\hat \phi_{I \setminus \vk},\hat V_J)\right. \\
		& + \sum_{\substack{K_1 \subset I\setminus \vk, \\ K_2 \subset J }} \left.\kappa(\hat \phi_\vq^{(s_\vk)},\hat \phi_{K_1}, \hat V_{K_2}) \kappa(\hat V_{\vk - \vq}^{(s_\vk)},\hat \phi_{K_1^c}, \hat V_{K_2^c}) \right].
	\end{split}
\end{equation}
We can use this equation to hierarchically expand any further cumulants. With every step of the expansion, a term $\hat V$ is added in the expansion introducing another small factor $\epsilon$. Furthermore, any cumulants $\kappa(\hat V_J)$ with more than 2 coefficients in the list $J$ give a zero contribution since the field $\hat V$ is Gaussian. The resonance condition will pick up exactly one term in the downward expansion. The upward cumulant hierarchy can be truncated due to the small coupling parameter $\epsilon$. For details we refer the reader to Ref.~\cite{Erdos2008} and references therein. We note that also the ergodic property for the random scattering system is well-known and referred to as the \emph{self-averaging property} in the mathematical physics literature.

\section{Decay of the energies away from the ring}
\label{sec:exponential_tail}
In this Section we justify the exponential decay of the temperatures $T(k)$ discussed in Sec.~\ref{sec:radial_distribution}. We assume that the temperatures (energies) of the modes $T(k)$ are isotropic i.e. depend only on the modulus $k$. For the random scattering case we refer the reader to Ref.~\cite{Heinonen2025}.

\subsection{Active Turbulence}
In order to analyze the average energies of the system we use Eq.~\eqref{eq:cumulant_time_stationary} to calculate the cumulant $\kappa (\hat \psi_\vk, \hat \psi_{-\vk}) = \langle \hat \psi_\vk \hat \psi_{-\vk} \rangle =: n_\vk$ giving
\begin{equation}
	2 D_\vk  n_\vk = -2 \sum\nolimits _{\vq_1} K_{\vk\vq_1} \Re[ \kappa(\hat \psi_{\vq_1}, \hat \psi_{\vk - \vq_1}, \hat \psi_{-\vk})].
	\label{eq:turbelence_second_order}
\end{equation}
Expanding the 3rd order cumulants using Eq.~\eqref{eq:cumulant_time_stationary} gives a 4th order cumulant and a product of two 2nd order cumulants. We write
\begin{widetext}
	\begin{equation}
		\begin{split}
			&D_I \kappa(\hat \psi_{\vq_1}, \hat \psi_{\vk - \vq_1}, \hat \psi_{-\vk}) = -\sum\nolimits_{\vq_2} K_{\vq_1\vq_2} \left[ 
			\kappa(\hat \psi_{\vq_2}, \hat \psi_{\vq_1 - \vq_2},\hat \psi_{\vk - \vq_1}, \hat \psi_{-\vk} ) + \kappa(\hat \psi_{\vq_2},\hat \psi_{-\vk}) \kappa(\hat \psi_{\vq_1 - \vq_2}, \hat \psi_{\vk - \vq_1} ) \right. \\
			&\left. +  \kappa(\hat \psi_{\vq_2},\hat \psi_{\vk - \vq_1}) \kappa(\hat \psi_{\vq_1 - \vq_2}, \hat \psi_{-\vk} )\right] 
			+ K_{\vk -\vq_1, \vq_2} \left[ \kappa(\hat \psi_{\vq_2}, \hat \psi_{\vk-\vq_1 - \vq_2}, \hat \psi_{\vq_1}, \hat \psi_{-\vk}) + \kappa(\hat \psi_{\vq_2}, \hat \psi_{\vq_1}) \kappa(\hat \psi_{\vk -\vq_1 - \vq_2}, \hat \psi_{-\vk}) \right. \\
			&+ \left. \kappa(\hat \psi_{\vq_2}, \hat \psi_{-\vk}) \kappa (\hat \psi_{\vk - \vq_1 - \vq_2}, \hat \psi_{\vq_1}) \right]
			+ K_{-\vk,\vq_2} \left[ \kappa(\hat \psi_{\vq_2}, \hat \psi_{-\vk - \vq_2}, \hat \psi_{\vq_1}, \hat \psi_{\vk - \vq_1} ) + \kappa(\hat \psi_{\vq_2}, \hat \psi_{\vq_1}) \kappa (\hat \psi_{-\vk - \vq_2}, \hat \psi_{\vk - \vq_1}) \right. \\
			& + \left. \kappa(\hat \psi_{\vq_2}, \hat \psi_{\vk - \vq_1}) \kappa (\hat \psi_{-\vk -\vq_2}, \hat \psi_{\vq_1}) 
			\right].
		\end{split}
	\end{equation}
\end{widetext}
We remind that the dissipation operator $D_I = D_{\vq_1} + D_{\vk - \vq_1} + D_{-\vk}$. 
The resonance condition \eqref{eq:resonance_2nd} gives up to second order cumulants 
\[ \begin{split}
	&D_I \kappa(\hat \psi_{\vq_1}, \hat \psi_{\vk - \vq_1}, \hat \psi_{-\vk}) =\\ &-(K_{\vq_1,\vk} + K _{\vq_1,\vq_1 - \vk}) n_{\vk} n_{\vk - \vq_1} \\
	&- (K_{\vk-\vq_1,-\vq_1} + K_{\vk - \vq_1, \vk}) n_{\vq_1} n_{\vk} \\
	&- (K_{-\vk,-\vq_1} + K_{-\vk,\vq_1-\vk}) n_{\vq_1} n_{\vk - \vq_1}.
\end{split} \]
Inserting this back in \eqref{eq:turbelence_second_order} and using a change of variables $\vq_1 \to \vk - \vq$ with the first row and $\vq_1 \to \vq$ for the rest of the rows gives 
\begin{equation}\label{eq:turbulence_second_order}
		D_{\vk} n_\vk = 2 \sum\nolimits_{\vq}\frac{K_{\vk\vq}}{D_I} n_\vq \left( K_{\vk\vq}  n_{\vk - \vq} 
		+ 2  K_{\vk - \vq,\vk} n_\vk \right)
\end{equation}
by using the symmetries of $K$ (see Eq.~\eqref{eq:A_operator} and below). We assume here that the 4th order cumulants can be neglected due to the near Gaussian property of the fields $\hat \psi$. 

We define the energy density as 
\begin{equation}
	\edens (\vk) = \frac{1}{8 \pi^2} A k^2 n_\vk
\end{equation}
and approximate the sum in Eq.~\eqref{eq:turbulence_second_order} as an integral giving
\begin{equation}\label{eq:energy_turbulence}
	\begin{split}
		D_\vk \edens (\vk) &= \int \diff \vq \frac{(\vk \wedge \vq)^2}{D_I k^2 q^2 |\vk -\vq|^2} (k^2 - 2 \vk \cdot \vq) \edens(\vq) \left [ \right . \\
		& \left . (k^2 - 2 \vk \cdot \vq) \edens(\vk - \vq) - 2 (k^2 - q^2) \edens(\vk) \right  ],
	\end{split}
\end{equation}
after substituting $K_{\vk\vq} = \frac 1 2 (\vk \wedge \vq) (|\vk - \vq|^2 - q^2)/k^2$.
Next we will examine Eq.~\eqref{eq:energy_turbulence} assuming one of the dominant interactions described in Fig.~\ref{fig:triad_interaction}. 

\textbf{Small} $\vq$:
This situation corresponds to Type \textbf{a} interaction in Fig.~\ref{fig:triad_interaction}. 
Eq.~\eqref{eq:energy_turbulence} is to lowest order in $\vq$
\begin{equation}\label{eq:type-a-interaction}
	D_\vk \edens(\vk) = 2 \int \diff \vq \frac{(\vk \wedge \hat \vq)^2}{D_I} \edens(\vq) \left(
	\edens(\vk - \vq) - \edens(\vk) \right),
\end{equation}
where $\hat \vq = \vq /q$. 
The extra factor 2 in front of the term $\edens_{\vk - \vq}$ comes from counting the symmetric term for which $|\vk -\vq|/|\vk| $ is small. In this case the second term becomes small so it only contributes when $|\vq|/|\vk|$ is small. The overall scaling of the coupling term is $\propto 1/D_I$. We note there that the limit $\lim_{\vq \to 0} \mathcal E (\vq)>0$.

\textbf{Equilateral triangle} $k\approx q \approx |\vk - \vq|$:
This corresponds to Type \textbf{b} interaction in Fig.~\ref{fig:triad_interaction}. Because of the scale constraint $k\approx q \approx |\vk - \vq|$, the triads form an an approximate equilateral triangle. 
We write $\vq = \tilde \vk - \vp$, where $\tilde \vk$ is the vector $\vk$ rotated 60 degrees and we assume $|\vp|/|\vk|$ to be small. Up to the lowest order in $\vp$ Eq.~\eqref{eq:energy_turbulence} becomes 
\begin{equation}\label{eq:type-b-interaction}
	\begin{split}
		D_\vk \edens(\vk) &=  2 \int \diff \vp \frac{(\vk \wedge \tilde \vk)^2}{D_{I} k^6} \vk \cdot \vp \edens(\tilde \vk - \vp) \left[ \vphantom{\edens_\vk \tilde \vk}\right . \\
		&\left . (\vk \cdot \vp) \edens(\vk - \tilde \vk + \vp) - 2 (\tilde \vk \cdot \vp) \edens(\vk) \right].
	\end{split}
\end{equation}
In this case $I = (\vk,\tilde \vk - \vp, \vk - \tilde \vk + \vp)$. The cross product gives $(\vk \wedge \tilde \vk)^2 = 3 k^4/4 $. Now the original sum over $\vq$ includes also a term, where $\tilde \vk$ is $\vk$ rotated $-60$ degrees. However, it suffices to notice that 
the overall contribution of this term scales as $\propto p^2 /D_I$ near $p = 0$.

Since in both of these cases all the modes in the triad are near the ring $k = \kcrit$, there is no difference in the operator $D_I$ or the magnitudes of the energies $\edens$ (latter condition is also verified numerically). Therefore we conclude that type \textbf{a} interactions dominate and the equilibrium value of $\edens(\vk)$ is approximately given by Eq.~\eqref{eq:type-a-interaction}. 

Next, we make the following assumptions:
\begin{enumerate}
	\item Energy density is isotropic i.e. $\edens(\vk) = \edens(k)$.
	\item The energy concentrated on the active ring at $k = \kcrit$ falls off rapidly away from the ring. Therefore type \textbf{a} interactions described by Eq.~\eqref{eq:type-a-interaction} dominate per the discussion above. 
	\item Energy density $\edens(q)$ is sharply peaked around $q = 0$.
\end{enumerate}
We write $\edens(\vq) = e^{g(\vq)}$. The last assumption allows for expanding the function $g$ up to a second order i.e. $g(\vq) \approx g(0) - \frac {\gamma^2} 2 q^2$. Here we use the fact that $\edens(\vq)$ has a maximum at $\vq = 0$ and it is isotropic. Notice that due to assumption 3 above, the parameter $\gamma \kcrit$ is large. 
We write
\begin{equation}
	\edens(\vq) \approx \edens_0 e^{-\frac {\gamma^2}{ 2 } q^2}.
\end{equation}

For any suitably well behaving function $f(\vq,\vk)$ we have
\begin{equation}\label{eq:qenergy-approximation}
	\int \diff \vq \edens(\vq) f(\vq,\vk) = \frac{1}{(2\pi)^2} \int \diff \vq \hat \edens(\vx) \hat f(\vx,\vk),
\end{equation}
where $\hat \edens$ and $\hat f$ are the Fourier transforms of $\edens$ and $f$ with respect to $\vq$. We have
\[ \hat \edens (\vx) = \edens_0 \frac{2\pi}{\gamma^2} e^{-\frac{x^2}{2 \gamma^2} }. \]
Because $\gamma$ is large, we can expand the exponential function giving
\[ \hat \edens (\vx) \approx \edens_0 \frac{2\pi}{\gamma^2} \left( 1 - \frac{x^2}{2 \gamma^2} \right). \]
Plugging this back in Eq.~\eqref{eq:qenergy-approximation} and using the inverse Fourier transform gives
\begin{equation}
	\int \diff \vq \edens(\vq) f(\vq,\vk)  \approx  \frac{2\pi \edens_0}{\gamma^2} \int \diff \vq \left (\delta(\vq) + \frac{\Delta}{2 \gamma^2} \delta(\vq) \right ) f(\vq,\vk),
\end{equation}
where $\delta$ is the Dirac delta and $\Delta$ is the Laplacian. Plugging this approximation back in Eq.~\eqref{eq:type-a-interaction} gives 
\begin{equation}
	\begin{split}
		D_\vk \edens (\vk) \approx &\frac{4 \pi \edens_0 }{ \gamma^2} \int \diff \vq \left (\delta(\vq) +  \frac{\Delta}{2 \gamma^2} \delta(\vq) \right ) \\
		&\times \frac{(\vk \wedge \vq)^2}{q^2 D_I} \left (\edens(\vk-\vq) - \edens(\vk)\right ).
	\end{split}
\end{equation}
We notice that the part without the Laplacian gives zero contribution due to the term $\edens(\vk - \vq) - \edens(\vk)$. Now,
\begin{equation}
	\frac{\gamma^4 D_\vk \edens (\vk)}{2\pi \edens_0 } \approx  \int \diff \vq \delta(\vq)  \Delta \left[ \frac{(\vk \wedge \vq)^2}{q^2 D_I} (\edens(\vk-\vq) - \edens(\vk)) \right].
\end{equation}
In the integrand we have a function multiplying the difference $\edens(\vk - \vq) - \edens(\vk)$. We can write the integral as
\[ \begin{split}
	&\int \diff \vq \delta(\vq) \left [ \left(\Delta \frac{(\vk \wedge \vq)^2}{q^2 D_I} \right) (\edens(\vk-\vq) - \edens(\vk)) \right .  \\
	&\left .+ 2\nabla \frac{(\vk \wedge \vq)^2}{q^2 D_I}\cdot \nabla \edens(\vk-\vq) + \frac{(\vk \wedge \vq)^2}{q^2 D_I} \Delta \edens(\vk -\vq)
	\right ] \\
	&=: I_1 + I_2 + I_3.
\end{split}
\]
In the following we use the identity
\begin{equation}
	\int \diff \vq \delta (\vq) f(q,\theta) = \lim_{q \to 0^+} \frac{1}{2\pi} \int_0^{2\pi} \diff \theta f(q,\theta).
\end{equation}
to evaluate integrals in polar coordinates. Using this to evaluate $I_3$ gives
\begin{equation}
	\begin{split}
		I_3 &= \Delta_\vk \edens(\vk) \frac{1}{2\pi}\int_0^{2\pi} \diff \theta \frac{k^2 \sin^2 \theta }{2 D(k) + D(0)} \\
		&= \frac{k^2}{4 D(k)} \edens''(k),
	\end{split}
\end{equation}
where, we define the \emph{explicitly} isotropic activation-dissipation operator $D(k) = D_\vk$.
Note here that the dissipation operator $D(0)=0$. 
The next integral can be evaluated in Cartesian coordinates with $\vk \cdot \vq =: k q_\parallel$ and $\vq \wedge \vk =: k q_\perp$. We notice that
\[ |\vk -\vq | = \sqrt{k^2 + q^2 - 2 k q_\parallel}  = k - q_\parallel + \mathcal{O} (q^2).\]
Therefore we have 
\[ \delta(\vq) \nabla_\vq \edens(\vk - \vq) = - \delta(\vq) \edens'(k) \hat \vk  \]
due to the isotropy of $\edens$. Now,
\[  
\begin{split}
	I_2 &= 2 \int \diff \vq \delta(\vq) \nabla \frac{(\vk \wedge \vq)^2}{q^2 D_I} \cdot \nabla \edens(\vk-\vq) \\
	&= -2 \edens'(k)k^2  \int \diff q_\perp \int \diff q_\parallel \delta(q_\perp) \delta (q_\parallel) \\
	& \parder{}{q_\parallel}\frac{ q_\perp^2}{q^2 (2 D(k) - q_\parallel D'(k))}
\end{split}
\]

We will evaluate this term in polar coordinates $(q_\parallel,q_\perp) = q (\cos \theta, \sin \theta)$ giving
\[   
\begin{split}
	I_2 =& -\frac{ k^2 \edens'(k)}{ D(k)} \int \diff q \frac{\delta(q)}{2\pi} \int_{0}^{2\pi} \diff \theta \\
	&\times \left ( \cos \theta \partial_q 
	- \frac{\sin \theta}{q} \partial_{\theta} \right )  \sin^2 \theta \left ( 1 + \frac{q \cos \theta D'(k)}{2 D(k)} \right )
\end{split}
\]
resulting in
\begin{equation}
	I_2 = - \frac{k^2 \edens'(k) D'(k)}{8 D(k)^2 }.
\end{equation}

For the last integral we have
\[ \edens(\vk- \vq) - \edens (\vk) \approx - q \cos \theta \edens'(k) + \mathcal{O}(q^2). \]
We write in polar coordinates 
\[ \begin{split}
	I_1 =& - \edens'(k) \int \diff q \frac{\delta(q)}{2\pi} \int \diff \theta q \cos \theta \\
	&\times \left ( \frac{1}{q} \partial_q (q \partial_q) + \frac{1}{q^2} \partial_\theta^2 \right ) \frac{(\vk \wedge \vq)^2 }{q^2 D_I}. 
\end{split} \]
We notice that taking the limit $q \to 0^+$ in the end accounts only for terms up to 1st order in $q$ in the term $1/D_I$. 
Thus we expand the kernel 
\[ \frac{(\vk \wedge \vq)^2 }{q^2 D_I} = \frac{k^2 \sin^2 \theta }{D(k) + D\left (\sqrt{k^2 + q^2 - 2 k q \cos \theta}\right ) + D(q)} \]
up to first order in $q$ giving
\[ \frac{(\vk \wedge \vq)^2 }{D_I} = \frac{\sin ^2 \theta }{2 D(k)} \left ( 1 +  \frac{q  \cos \theta  D'(k) }{2 D(k)} \right ) + \mathcal{O}(q^2).\]
We can use integration by parts for the part with the derivative $\partial_\theta^2$. We notice that the $q$ and $\theta$ derivatives cancel out for the first order term. The only term left is the 0th order term:
\[ I_1 = \frac{\edens'(k) k^2}{2\pi} \lim_{q \to 0^+} \int_{0}^{2\pi} \diff \theta \frac{\cos \theta \sin ^2 \theta }{2 D(k) q} = 0. \]

Combining $I_2$ and $I_3$ gives
\begin{equation}\label{eq:AT-energy-diff-eq}
	C^2 \frac{D(k)^2}{k^2}  \edens(k) = \dfrac{k_\text{m}^4}{\gamma^4} \left ( \edens''(k) - \frac{D'(k)}{2 D(k)} \edens'(k)
	\right ),
\end{equation}
where $C^2 = 2 k_\text{m}^4/(\pi \edens_0)$ and $k_\text{m} \approx \kcrit$ maximizes $D(k)$.
This equation can be solved perturbatively with the assumption that the parameter $\gamma/k_\text{m}$ is large. We write 
\begin{equation}
	\edens(k) = \exp \left (\f 1 \delta \sum_{j=0}^{\infty} \delta^j Q_j(k) \right ),
\end{equation}
where $\delta = k_\text{m}^2/\gamma^2$, which is assumed to be a small parameter. To the lowest order we have 
\begin{equation}
	C^2 \frac{D(k)^2}{k^2} = Q_0'(k)^2,
\end{equation}
which has two solutions 
\begin{equation}
	Q_0'(k) = \pm C \frac{D(k)}{k}.
\end{equation}
We can expand $D(k)$ around the maximum $k_{\text{m}}$ and integrate from $k_{\text{m}}$ to $k$. To the lowest order this gives
\begin{equation}
	Q_0(k) - Q_0(k_{\text{m}}) = \pm C D(k_{\text{m}}) \left ( \frac{k}{k_{\text{m}}} - 1 \right ),
\end{equation}
which in terms of energy is
\begin{equation}\label{eq:AT-energy-decay}
	\edens(k) \approx \edens(\kmax) \exp \left ( \pm  D(\kmax) \sqrt{\frac{2 \gamma^4}{\pi \edens_0}} \left ( \dfrac{k}{\kmax} -1 \right ) \right ).
\end{equation}
We assume that the sign change happens at $k = \kmax$ giving exponential decay for both $k < \kmax$ and $k > \kmax$. 

In this calculation we have assumed that $D_I = D(k) + D(|\vk - \vq|) + D(q) \neq 0$. When $k \approx \kmax$ the kernel $\edens(q)$ goes to zero fast enough ensuring that contribution of the singular integrals over the zero level set of $D_I$ are negligible. However, for $k$ s.t. $D(k) \approx 0$ this is not true and therefore this result cannot be used for all $k$. Note also that the singularity at $k = \kmax$ in Eq.~\eqref{eq:AT-energy-decay} is artificial for finite $\delta$. Instead, we expect the exponent to be continuous and therefore displaying Gaussian behavior around $k = \kmax$. 
Eq.~\eqref{eq:AT-energy-decay} is still expected to give a good description of the decay of the radial energy away from $\kmax$ keeping in mind that proper description of the asymptotic behavior with $|k - \kmax| \gg \kmax $ requires an analysis including Type \textbf{b} interactions.


\section{Range of validity}\label{sec:breakdown}
Universality in equilibrium systems arises from symmetries between the microscopic states of the system. For example, the microstates of a microcanonical system are assigned a uniform probability distribution -- a state that is achieved through some notion of ergodicity. For canonical systems it is assumed that the entropy of the reservoir is much larger than the entropy of the canonical subsystem. Universality also arises in statistical field theories in the form of conformal symmetries sufficiently close to thermodynamic critical points. 
Similarly, it is interesting to explore under which conditions deviations from the universal energy distribution Eq.~\eqref{eq:pdf_large_energy} arise in non-equilibrium systems. 

\par
\begin{figure*}[]
	\centering
	\includegraphics[width=2\columnwidth]{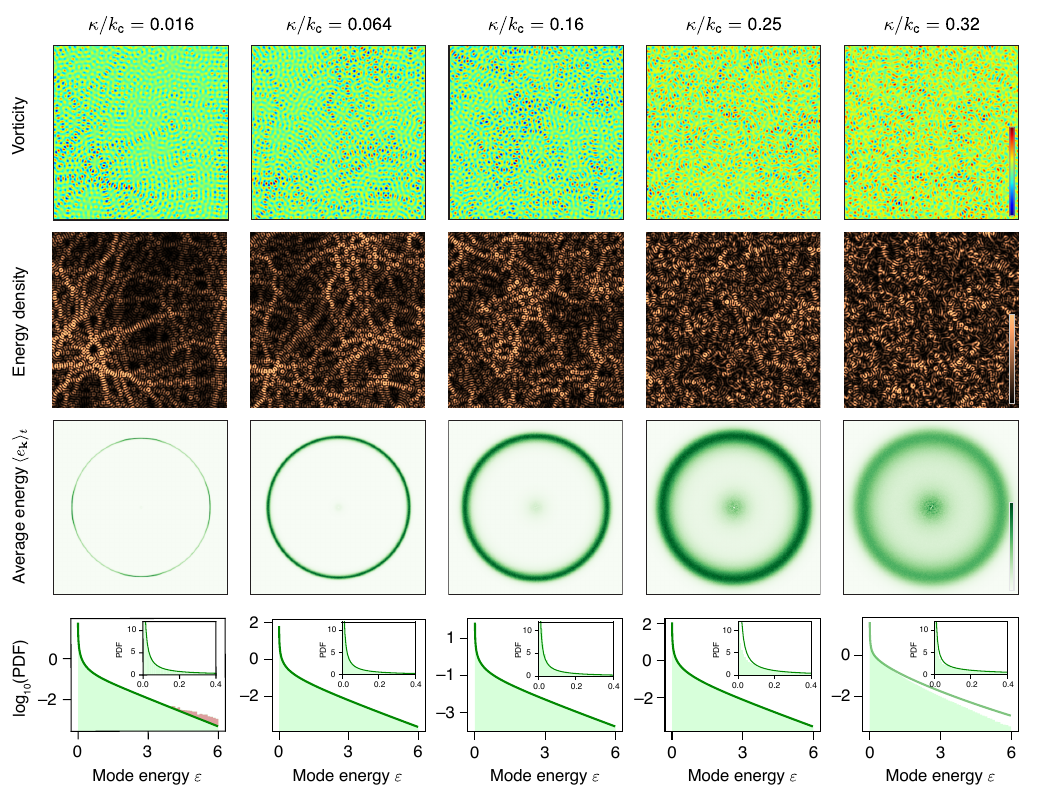}
	\caption{First and second rows show the vorticity fields and real space energy densities for a representative time for varying active ring width in the active turbulence simulations. The third row shows the spectral energy density while the probability density functions are shown on the last row. We use the orange color to emphasize the deviation from the predicted behavior at high energies. Apart from the ring width parameter $ \kappa $, the other parameters are the same as the parameters for the active turbulence simulation in Main Text {(for results in the Main Text, $\kappa/\kcrit=0.3/\pi\approx 0.096$)}.
	}
	\label{fig:breakdown}
\end{figure*}

We tested the applicability of the theory presented in this work by simulating the active turbulence system with varying width of the active ring. Fig.~\ref{fig:breakdown} shows a gradual breakdown of the predicted energy PDF for very narrow and large rings. For very narrow rings we see that the energy on the $k$ ring is not isotropic. Due to the definition of the energy exchange operator $ D $ for the active turbulence, the dissipation of the modes close to $ k=0 $ increases as the ring is made narrower. For extremely narrow rings just the four modes adjacent to the mode $ k=0 $ are occupied. Since the mode coupling between the modes on the ring is facilitated by these large wavelength modes (see Fig.~\ref{fig:triad_interaction}), this makes the coupling anisotropic. Consequently, we observe accumulation of energy near the modes $ k_x=0 $  and $ k_y = 0 $. We tested this hypothesis by increasing the system size and affirmed that indeed the energy will be distributed isotropically with equally thin ring. One should note that this is not a numerical artifact. Instead, this is a physical effect of the finite domain size. 

Fig.~\ref{fig:breakdown} also shows a deviation from the theoretical energy distribution for sufficiently large widths of the active ring. In this case it is not expected that the radial energy distribution on the different $ \vec{k} $ modes follows the shape introduced in Sec.~\ref{sec:radial_distribution}. The universality of the distribution is a consequence of the narrowness of the ring and it is expected that the distributions lose universality when the length scale selection mechanism is relaxed.

\begin{figure*}[]
	\centering
	\includegraphics[width=2\columnwidth]{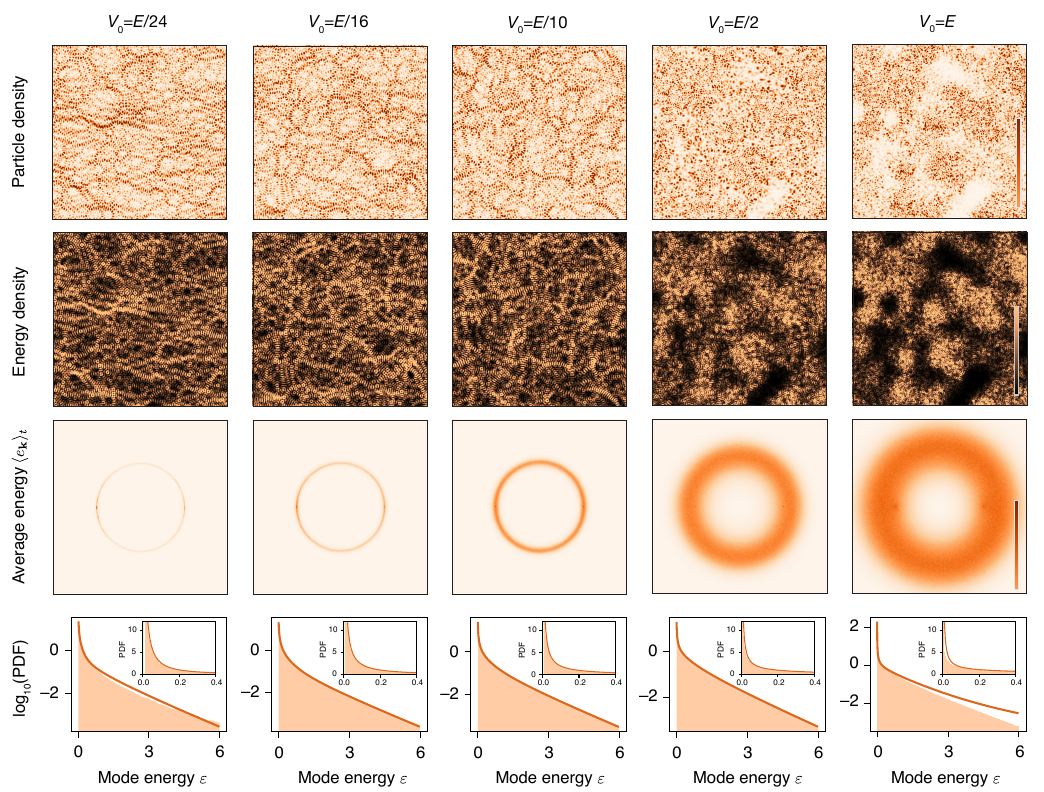}
	\caption{First and second rows show the particle and real space energy densities for a representative time for varying strength of the scalar potential $ V $ in the random scattering simulations. Here $ V_0 $ is the average of the total potential energy contribution {$ \sqrt{\langle V(x)^2 \rangle} $} and {$ E = \kcrit^2/2 $} is the initial kinetic energy of the incident plane wave. The third row shows the spectral energy density while the probability density functions are shown on the last row. Other parameters are as described in Main Text. 
	}
	\label{fig:breakdown_quantum}
\end{figure*}

Fig.~\ref{fig:breakdown_quantum} shows a similar test for the random scattering case. We can control the ring width by changing the magnitude of the smooth background potential while keeping other properties the same. We see that for weak enough potentials the energy will not be distributed isotropically on the ring. We see large peaks at the incidence wave $ \vec{k} $ number $\vec{k} = (k_\text{c},0)  $ and a corresponding backscattering peak at $ \vec{k} = (-k_\text{c},0) $. The background potential is too weak to scatter the incident wave isotropically and the orientational order of the initial configuration is retained. This will cause deviations in the overall energy statistics. Note that the reason the system remains anisotropic is unique to the random scattering system. For strong enough background potentials the ring becomes extremely wide and the potential localizes quantum particles in energy wells with random shapes. We see again that the radial energy distribution looses universality resulting in a breakdown of the theory. 

Secs.~\ref{sec:cumulants} and \ref{sec:exponential_tail} give some insights to the mechanisms leading to universality. The theory does not depend on the exact statistics of the quantum potential $V$ or the precise shape of the dissipation-activation operator $D$ as long as the length-scale selection is maintained and the coupling between the Fourier modes is weak. 
We note that the theory works for a wide range of different ring widths and it seems that the underlying assumptions can be relaxed significantly until a deviation in the overall energy statistics is observed. This suggests that the theory has a wide range of applicability.

\clearpage
\bibliography{draft}

\end{document}